\documentclass[apj]{emulateapj}
\shorttitle{CMB Cold Spot in the {\it Planck} light}
\shortauthors{M.~ Farhang, S. M. S. Movahed}
\usepackage{color}
\usepackage[colorlinks=true,
            linkcolor=black,
            urlcolor=black,
            citecolor=blue, breaklinks=true]{hyperref}
            
\usepackage{url}
\usepackage{selinput}
\SelectInputMappings{%
	agrave={à},
	ccedilla={ç},
	Euro={¤}
}
\usepackage{amsmath}
\usepackage{amssymb}
\usepackage{gensymb}
\usepackage{multirow}
\usepackage{graphics}
\usepackage{graphicx}
\usepackage{natbib}
\usepackage{psfig}
\input{epsf}
\usepackage{ae,aecompl}
\usepackage{booktabs}

\newcommand{\be}{\begin{equation}}
\newcommand{\ee}{\end{equation}}

\begin{document}
\title{CMB Cold Spot in the {\it Planck} light}

\author{M. Farhang, S. M. S. Movahed}
\affil{\scriptsize
$^{1}${Department of Physics, Shahid Beheshti University, 1983969411,  Tehran Iran}\\
}

\email{Corresponding author email: m\_farhang@sbu.ac.ir}

\begin{abstract}
The Cold Spot, with an unusually cold region  surrounded by a hot ring, is a statistically significant anomaly in the Cosmic Microwave Background (CMB) sky. In this work we assess whether different sets of multiple subvoids based on the 2dF-VST ATLAS Cold Spot galaxy redshift survey or a collapsing cosmic texture could have produced such an anomaly through a simultaneous search for their gravitational redshift and lensing signatures on the {\it Planck} CMB temperature anisotropies.
We use patches with radii $R=10^\circ$ and $R=20^\circ$ to account for the inner cold region as well as the outer hot  ring. 
As the void model, we explore two sets of $\Lambda$LTB templates characterized by different values of the model's free parameters, and a top-hat void template.
We detect higher-than-expected gravitational redshift amplitudes for the first two sets,   $\mathcal{A}_{\rm rs}=5.4\pm 1.4$ and $\mathcal{A}_{\rm rs}=14.4\pm 3.8$,  and lower than expected for the top-hat model, $\mathcal{A}_{\rm rs}=0.3 \pm 0.1$. 
The amplitudes for the lensing imprint are consistent with zero for all these subvoid models.
%
The estimated amplitude for the texture imprint from the gravitational redshift measurement implies 
the energy scale of the texture, parametrized by $\epsilon$, to be $\epsilon= (7.6\pm2.0)\times 10^{-5}$, with no detection of the lensing trace.
 We note that the deviation of the subvoid amplitudes from unity and the inability of the texture and some of the void profiles to reproduce the hot ring  indicate theoretical insufficiencies, either in the construction of the model or in the assumed gravitational and cosmological framework leading to the imprints for the  structures. 
 \end{abstract}

\section{Introduction}
The Cold Spot (CS) in the Cosmic Microwave Background (CMB) map is an unusually cold and large region, located at $(l, b)\sim(208^{\circ},-57^{\circ})$ with the mean temperature decrement $\Delta T\approx -150$  $\mu$K in a  $5^\circ$ radius,  and is among the robust large-scale CMB anomalies.
 It was first detected in the WMAP data through the Gaussianity test of CMB anisotropies using Spherical Mexican Hat Wavelets with $\approx (2 \sim 3) \sigma$ significance \citep{vie04} and its significance and properties were further explored through other techniques \citep{larson2004hot,cru05,cru06,cru07a,vie10}.
 The existence and anomalous nature of the CS in the $\Lambda$CDM model was  subsequently confirmed by  {\it Planck} observations \citep{pl13,pl15}. 
Although originally detected due to its unusual coldness, it is  argued that 
the anomalous nature of the CS is in fact due to its peculiar template of a central cold region surrounded by a hot ring, subtending to the radius of $R\sim 15^\circ$ \citep{zha09,nad14,kov17} 

The chances
that Gaussian and isotropic initial conditions, as predicted by standard inflationary scenarios (see, e.g., \citealt{lid00}), have led to such a
large and relatively cold region are low, at $0.5\%$ level \citep{cru05,cru06,cru07}. See  \cite{zha09} for a different viewpoint. 
The roles of systematics  and foreground contamination are shown to be negligible in producing the CS \citep{cru06a,cru04}.
There have been therefore speculations of the CS being
produced by secondary sources of anisotropy such as a supervoid or a
cosmic texture.
  \cite{ino06,ino07} showed that  a huge underdense region with density contrast $\delta_{\rm V} \sim -0.3$ and the comoving radius  $R_{\rm V}\sim 200-300 {\rm Mpc}/h$ located at $z_{\rm V} \approx 1$ could lead to such a pattern of anisotropy. 
   A completely empty void, i.e., with $\delta_{\rm V} \sim -1$, would require a smaller radius $R_{\rm V} \sim 120 {\rm Mpc}/h$ to agree with observations \citep{rud07}.
  However, $\Lambda$CDM-based simulations suggest typical void sizes of $R_{\rm V} \sim 10 {\rm Mpc}/h$ (with $\delta_{\rm V} \sim -0.8$), rendering  such huge voids  unlikely to find \citep{pat06,hoy04,col05,pla08,mac17}.
 Also lack of high redshift anomalies in the redshift distribution of galaxies rules out the existence of a high redshift supervoid in the direction of the CS anomaly \citep{smi08, gra09,bre10,pat06,hoy04,mac17}. 

 On the other hand, there is growing observational evidence supporting the existence of a shallow but large supervoid at low redshift, in the constellation Eridanus, aligned with the CS line of sight \citep{rud07,smi08,gra09,bre10,sza14}.
 Galaxy catalogues are consistent with a large supervoid with the radius $R_V=220\pm 50$ Mpc/$h$, and underdensity $\delta_V=-0.14\pm 0.04$ centered at $z_{\rm V}=0.22 \pm 0.03$ \citep{sza14, kov15,cou17}. However, the actual shape of this rare matter fluctuation is not fully clear yet \citep{mac17,kov15}.  %
  The secondary CMB temperature decrement due to such a supervoid is 
 estimated to be $\sim 20$ $\mu$K, much smaller than the observed CS temperature, and the predicted angular profile of CMB temperature anisotropy generated by certain models of  supervoids in the $\Lambda$CDM cosmology differs from the observed CMB cold spot \citep{fin14,nad14, zib14,mar16,sza14}. 
 \cite{mac17} explored the substructure of the CS region in the 2dF-VST ATLAS Cold Spot galaxy redshift survey and reported four subvoids in the redshift range of $0.14-0.42$. They also concluded that the sum of the induced anisotropies by these voids is too low to explain the observed CMB CS. 
 
 However, recent studies have claimed that there is probably a factor of $5-10$ mismatch between the simulation-based and observed  temperature decrements from various supervoids \citep{kov17}.  
  This excess signal offers a challenging opportunity for exploring beyond $\Lambda$CDM theories (e.g., \citealt{rud07,pap10a,pap10b,fle13,hot14}).
  It has  been argued that supervoids could lead to enhanced signature in cosmological models other than $\Lambda$CDM (see, e.g., \citealt{bec18}). For other non-cosmological  suggestions as possible ways out of this mismatch see \citet{fle13,kovb15}.\\

 An alternative to voids,  first proposed by \cite{cru07},  is a collapsing cosmic texture. Collapsing textures could cause a CS on the CMB sky through gravitational interaction with the photons passing nearby.
 We here note that there is argument in the literature against collapsing cosmic textures as viable sources of the CS anomaly.  \cite{fen12} use a hierarchical Bayesian analysis to search for the texture model in WMAP7 data where  selection biases are automatically accounted for. They found no statistical preference for the texture model compared to the standard model with no textures (see also \citealt{pei14}). 
Further independent probes in the search of collapsing cosmic textures are necessary to assess the reality of their existence. Textures could lens the CMB E-mode anisotropies and turn them partially to B-mode. This B-mode signal could serve as a texture probe \citep{gar11}. 
Lensing by texture would also leave an imprint on the 21 cm signal, if given long dedicated observation time \citep{kov13}.
In spite of the above arguments against textures, we still include a collapsing texture
among the candidates for the CS origin. Given the importance of the {\it Planck} dataset and the history of the study of the texture as a candidate, we find it worthwhile to see what {\it Planck} has to say about it, irrespective of, and unbiased by, other datasets, simply as a candidate along with more viable candidates.

%
More specifically,  CMB photons passing through or nearby these structures, on their way to reach us, experience gravitational redshift and lensing. 
We simultaneously measure, for the first time, the amplitudes of these two imprints for  voids and a cosmic texture as observed by {\it Planck} and assess their consistency. 
If consistent, these measurements would imply the viability of the assumptions made in the template construction.
Their inconsistency, on the other hand, may call for a different parameterization of the templates or different parameter values, or even more severely,  challenge the role of these structure in generating the CS. 

The rest of this paper is organized as follows.
In Section~\ref{candid} we introduce two sets of well-motivated candidates, 
voids and a collapsing cosmic texture, as possible sources of the CS and discuss their gravitational redshift and lensing imprints on the temperature of CMB photons.
%
The mathematical framework for the analysis of these imprints is explained in Section~\ref{anal}, and the results are presented in Section~\ref{res}. We conclude in Section~\ref{concl}.
Whenever needed, the standard $\Lambda$CDM cosmology, consistent with the {\it Planck} 2018 data \citep{pl18}, is assumed throughout.

\section{Candidates}\label{candid}
In this section we introduce two sets of physically motivated candidates as possible origins of the CMB CS, i.e., voids in Section~\ref{sec:void} and a collapsing cosmic texture  in Section~\ref{sec:texture}. We discuss the gravitational redshift and lensing of CMB photons as they pass through or close to these intervening structures.
\subsection{Multiple Subvoids}\label{sec:void}
An expanding local void can produce gravitational redshift and thus temperature decrements in CMB photons passing through it, 
 through the combination of the linear Integrated Sachs-Wolfe (ISW) \citep{sac67} and the nonlinear Rees-Sciama effects \citep{ree68}.
 The linear effect is shown to substantially dominate the second order nonlinear contribution \citep{cai10, nad14}. The nonlinear effect is not therefore considered here.
This ISW imprint is sensitive to the expansion history of the Universe at late times and is therefor considered as a probe of dark energy \citep{peebles1984tests,hu1996acoustic, afshordi2004integrated,acquaviva2006dark,schafer2008integrated,carbone2013maps,amendola2008measuring,ade2016planck,mostaghel2018integrated}.
 
 In this section, we introduce two sets of void models. 
 In Section~\ref{sec:mackenzie} we use as our void templates the multiple subvoids reported by \cite{mac17} found through mapping the Cold Spot region. 
 These voids are modeled as (basic and modified) $\Lambda$LTB spherical underdensities \citep{fin14}. 
 For reasons to be discussed, we also use a different void profile in Section~\ref{sec:martinez}, described by a sharply compensated top-hat radial template \citep{martinez1990anisotropies,mar90,cru08}. 
\subsubsection{$\Lambda$LTB Subvoids}\label{sec:mackenzie}
The substructure of the core of the CMB cold spot has been recently explored based on the 2dF-VST ATLAS Cold Spot galaxy redshift survey \citep{mac17}.
They reported voids at four redshifts with various sizes and underdensities. 
We use these subvoids as our main set of void candidates and explore whether the CMB CS could be interpreted as the combined ISW imprints of these voids.

We follow the void underdensity template used in \cite{mac17}. It was originally proposed by \cite{fin14} in the context of CS as an underdensity in the  $\Lambda$LTB metric \citep{gar08}
\begin{equation}\label{eq:ltm}
ds^2=-dt^2+\frac{A'(r,t)^2}{1-k(r)}dr^2+A(r,t)^2d\Omega^2
\end{equation}
with the following spatial curvature
\begin{equation}\label{eq:kltb}
k(r)=k_0r^2\exp\left( -\frac{r^2}{R_{\rm V}^2}\right),
\end{equation}
where $R_{\rm V}$ is the radius of the spherically symmetric underdensity. With the FRW metric, $A(r,t)=a(t)r$ where $a(t)$ is scale factor.
This void model is shown to be well approximated by a linear potential perturbation $\Phi$ in a flat FRW background, where 
\begin{equation}\label{eq:phi}
\Phi(\tau,r)=\Phi_0\exp\left( -\frac{r^2}{R_{\rm V}^2}\right) {}_2F_1\left[1,\frac{1}{3},\frac{11}{6},\frac{-\Omega_{\Lambda}}{\Omega_{\bf m}(1+z)^3}\right].
\end{equation}
Here  ${}_2F_1$ is a  hypergeometric function
and $\tau$ is the conformal time in the FRW metric. On can use the Poisson equation to get the density profile for the potential of Eq.~\ref{eq:phi}
\begin{equation}
\delta(r,z_{\rm V})=\delta_{\rm V} g(z_{\rm V})\left(1-\frac{2}{3}\frac{r^2}{R_{\rm V}^2}\right)\exp\left(-\frac{r^2}{R_{\rm V}^2}\right), \label{eq:delta_subv}
\end{equation}
where $\delta_{\rm V}$ and  $z_{\rm V}$ are the void underdensity and redshift.
Also $r$ is the distance from the void center and $g(z)$ is the growth factor at redshift $z$.
We also have 
\begin{equation}
\Phi_0=\frac{\Omega_{\rm{m}}}{4}\frac{H_0^2R_{\rm V}^2\delta_{\rm V}}{{}_2F_1\left[1,\frac{1}{3},\frac{11}{6},\frac{-\Omega_{\Lambda}}{\Omega_{\bf m}}\right]}.
\end{equation}
The induced ISW  temperature anisotropies in CMB photons by this density contrast would be
%
\begin{eqnarray} \label{eq:ltb}
&&\delta T_{\rm rs}(\theta) \approx  \frac{3\sqrt{\pi}}{22} \delta_{\rm V}\frac{H(z_{\rm V})\Omega_\Lambda ~{}_2F_1\left[2,\frac{4}{3},\frac{17}{6},\frac{-\Omega_{\Lambda}}{\Omega_{\bf m}(1+z_{\bf V})^3}\right]}{H_0(1+z_{\rm V})^4~{}_2F_1\left[1,\frac{1}{3},\frac{11}{6},\frac{-\Omega_{\Lambda}}{\Omega_{\bf m}}\right]}  \nonumber \\ 
 &&\times \left(H_0R_{\rm V}/c\right)^3\left(1+{\rm erf}\left[\frac{cz_{\rm V}}{H(z_{\rm V})R_{\rm V}} \right]\right)\exp\left(-\frac{ \theta^2}{\theta_{\rm V}^2} \right)
\end{eqnarray}
where $\delta T\equiv \Delta T/T_{\rm CMB}$
 and the subscript rs labels the redshifted signal.   
We stress that the main assumption in this model is the Guassian profile of the spatial curvature in the Eq.~\ref{eq:kltb}. The induced anisotropy is shown to be equivalent whether calculated in an exact $\Lambda$TBM framework or approximated as a perturbation in a FRW universe \citep{nad14}.

%
\begin{figure}
\begin{center}
\includegraphics[scale=0.4]{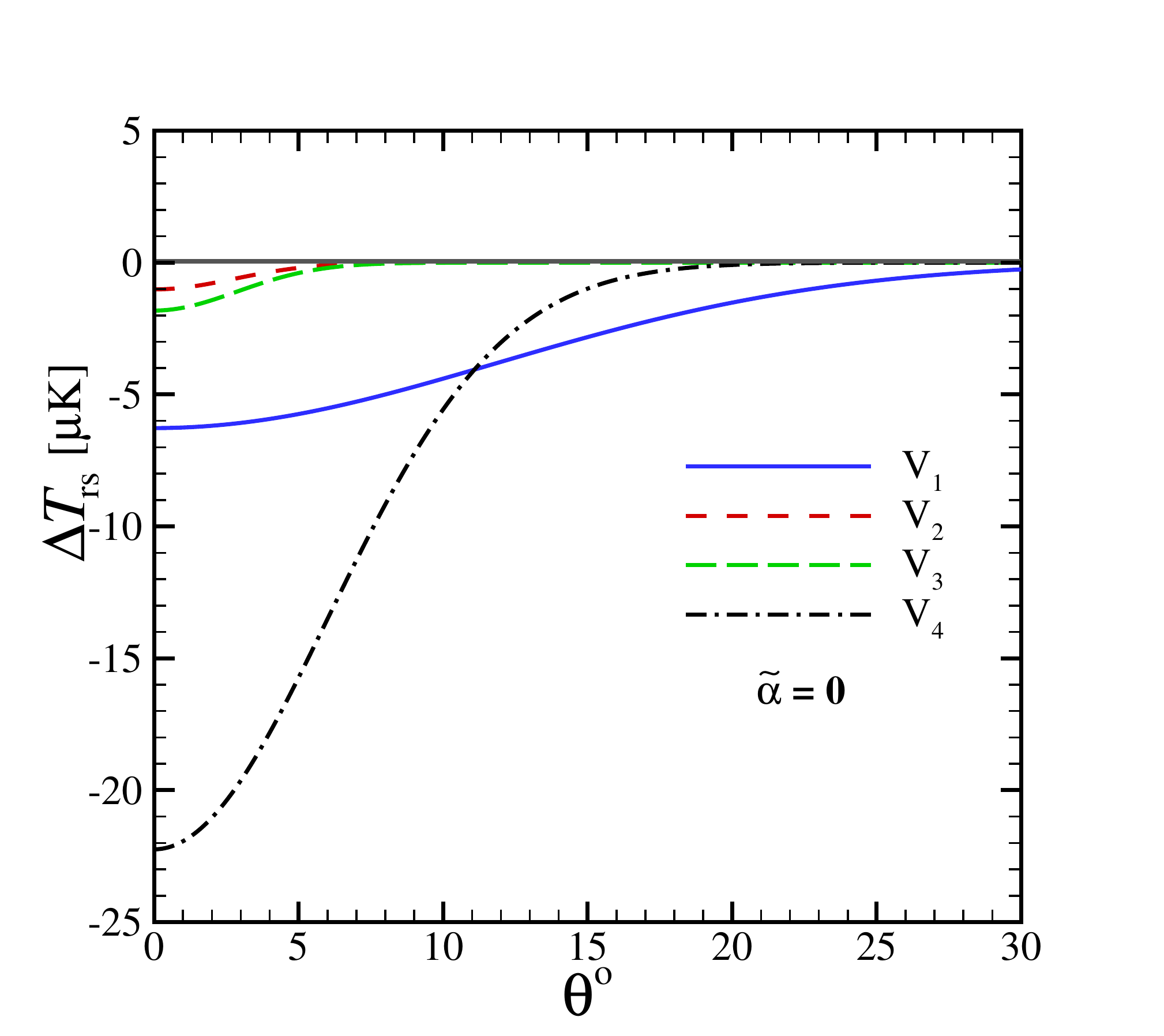}
\includegraphics[scale=0.4]{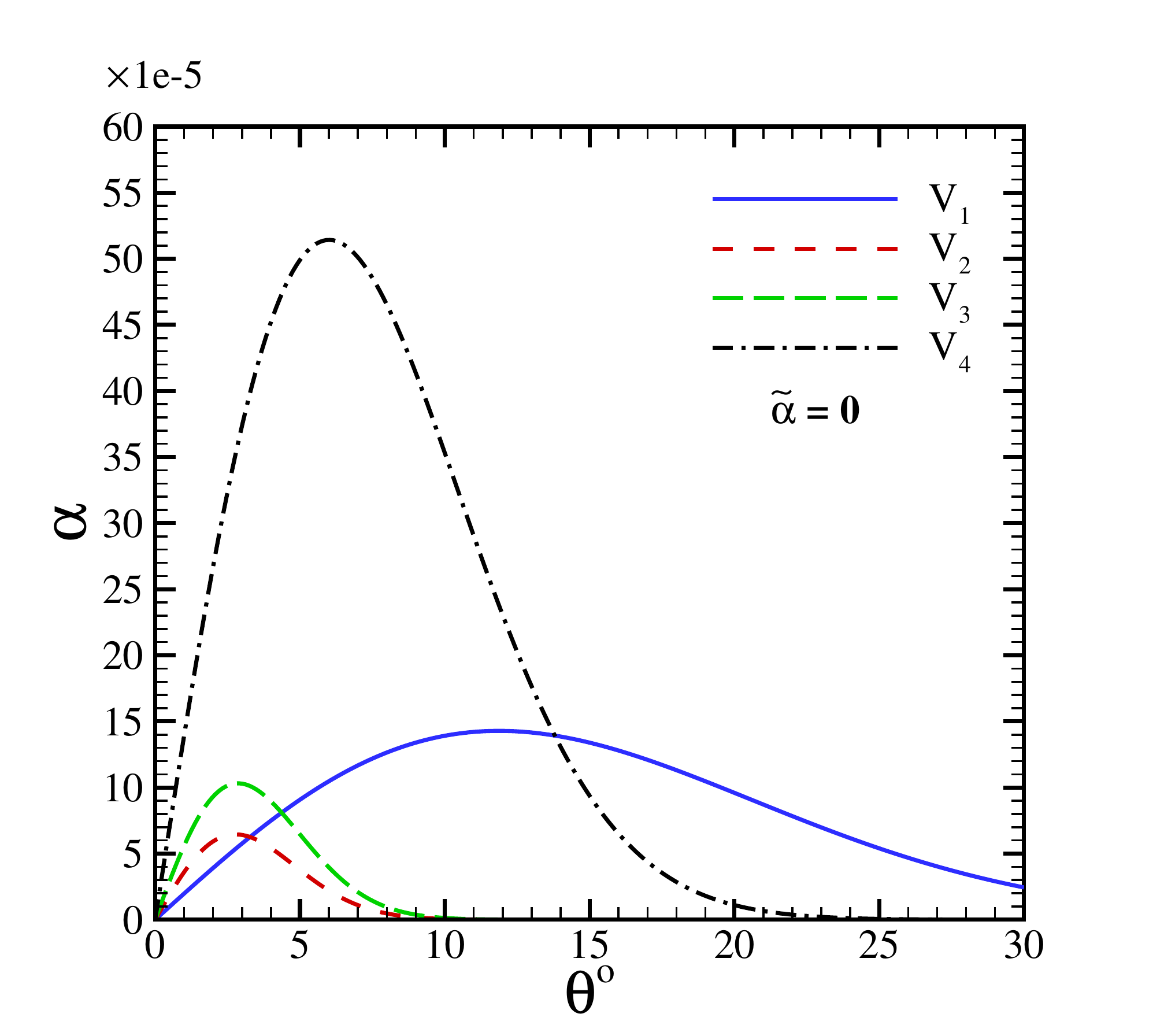}
\end{center}
\begin{center}
\caption{Up: The radial profiles of temperature anisotropies due to the gravitational redshift (here only ISW) of the basic $\Lambda$LTB subvoids for $\tilde\alpha=0$ characterized by the parameters of  Table~\ref{table:subv}. Bottom: The corresponding radial profiles of the lensing deflection angle.}\label{fig1}
\end{center}
\end{figure}


%
 Besides the gravitational redshift,  photons that pass through the void also experience gravitational lensing.
 The lensed temperature field, $\delta \tilde{T}$, for small radial deflection angle is expressed as
\begin{equation}\label{eq:lens}
\delta {\tilde T}(\theta)=\delta T(\theta+\alpha(\theta))\simeq \delta T(\theta)+ \delta T_{\rm ln}(\theta),
\end{equation}
where  $\delta T$  is the unlensed map, $\delta T_{\rm ln}\equiv \alpha(\theta)\partial \delta T(\theta)/\partial \theta$ and 
$\alpha(\theta)$ is the template for the lensing deflection angle. The subscript ln labels the gravitational lensing effect. 
The deflection angle induced by the void of Eq.~\ref{eq:delta_subv} is given by
\begin{eqnarray}
\alpha(\theta)&\approx&-\delta_{\rm V}\frac{D_{\rm LS}D_{\rm L}}{D_{\rm S}}\frac{2R_{\rm V}H_0^2}{c^2}\frac{\Omega_m}{{}_2F_1\left[1,\frac{1}{3},\frac{11}{6},\frac{-\Omega_{\Lambda}}{\Omega_{\bf m}}\right]}  \nonumber \\ 
&& \times  { {}_2F_1\left[1,\frac{1}{3},\frac{11}{6},\frac{-\Omega_{\Lambda}}{\Omega_{\bf m}(1+z_{\rm V})^3}\right]\theta \exp\left(-\frac{\theta^2}{\theta_{\rm V}^2}\right) }.\label{eq:alfa:ltbmv}
\end{eqnarray}
where $D_{\rm L}$ and $D_{\rm S}$ are the observer's comoving distances to the lens (here, the center of the void) and the source (here, the last scattering surface),
and $D_{\rm LS}$ represents the source distance to the lens. 
 Table~\ref{table:subv} presents the subvoid characteristics as used in this work, based on Table 3 of \cite{mac17}, and Figure~\ref{fig1} illustrates the corresponding radial profiles for the induced ISW anisotropies (up) and the lensing deflection angle (bottom).
 We refer to these basic $\Lambda$LTB subvoids as $V_1$-$V_4$.
It is evident from the figure and also expected from Eq.~\ref{eq:ltb} that the $\Lambda$LTB voids can not produce hot rings, or even more generally, any temperature increments. 
\cite{fin14}  therefore modified the basic $\Lambda$LTB template to allow for the production of a hot ring. We also explore the implications of considering this new set of templates, here called modified $\Lambda$LTB, as candidates for CS source. 
%
\begin{table}
\centering
\caption{Characteristics of the subvoid candidates considered in Section~\ref{sec:void}, described by the triplets  $(R_{\rm V}, z_{\rm V}, \delta_{\rm V})$ representing the  comoving radius, redshift and underdensity of the subvoids, respectively. The values are taken from Table 3 of \cite{mac17}.}
 \begin{tabular}{ccccccc}
\hline\hline 
 & subvoid 1 & subvoid 2  & subvoid 3 & subvoid 4 \\ \hline
 $R_{\rm V}$   (Mpc/$h$)   &$119$  & $50$&$59$  &$168$  \\ \hline
  $z_{\rm V}$  &$ 0.14$ &  $ 0.26$ & $0.30$  & $ 0.42$          \\\hline
   $\delta_{\rm V}$  &$ -0.34$ &  $ -0.87$ & $-1.00$ & $ -0.62$ \\
  \hline
\end{tabular}\label{table:subv}
\end{table}
\begin{figure}
\begin{center}
\includegraphics[scale=0.4]{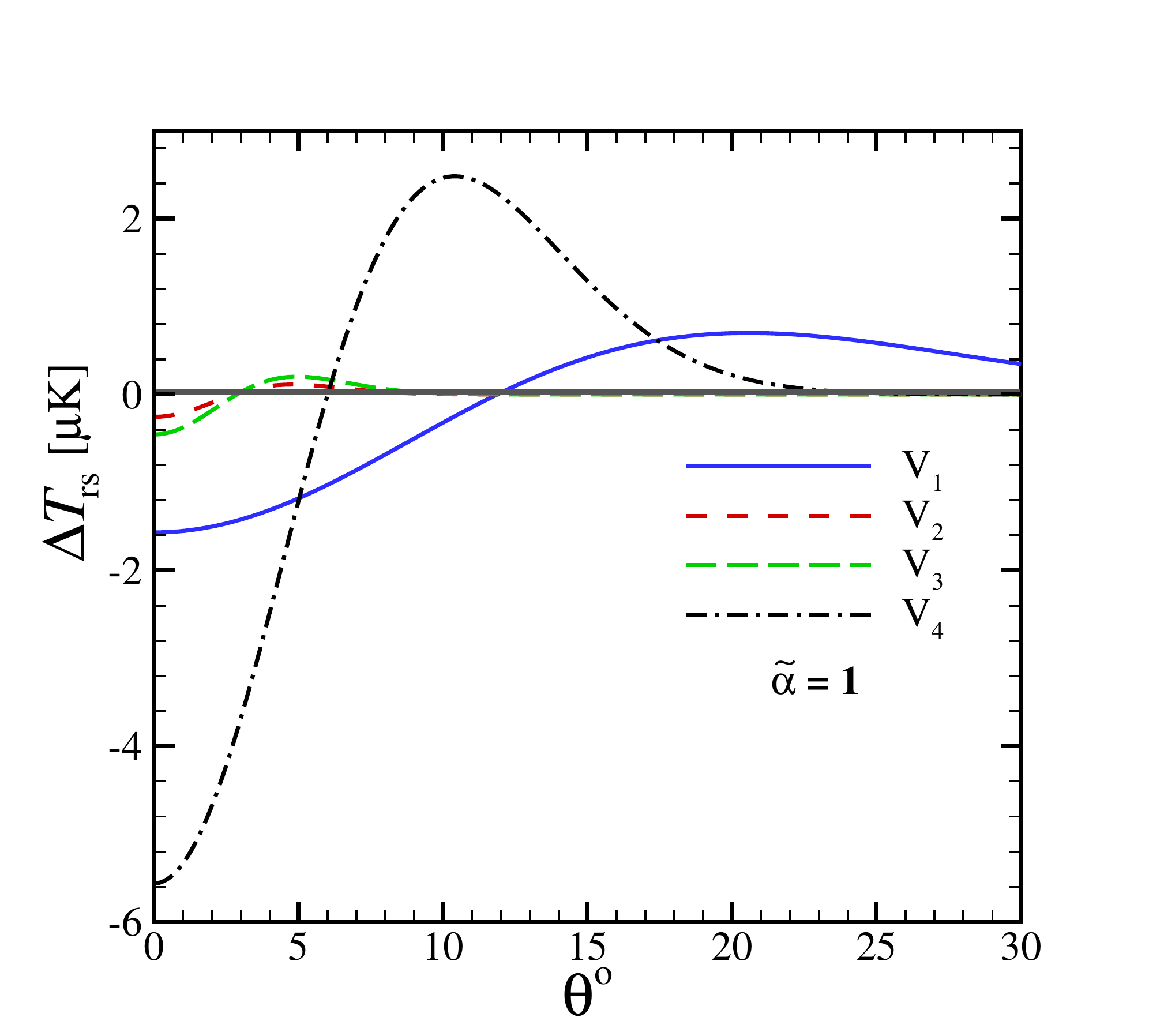}
\includegraphics[scale=0.4]{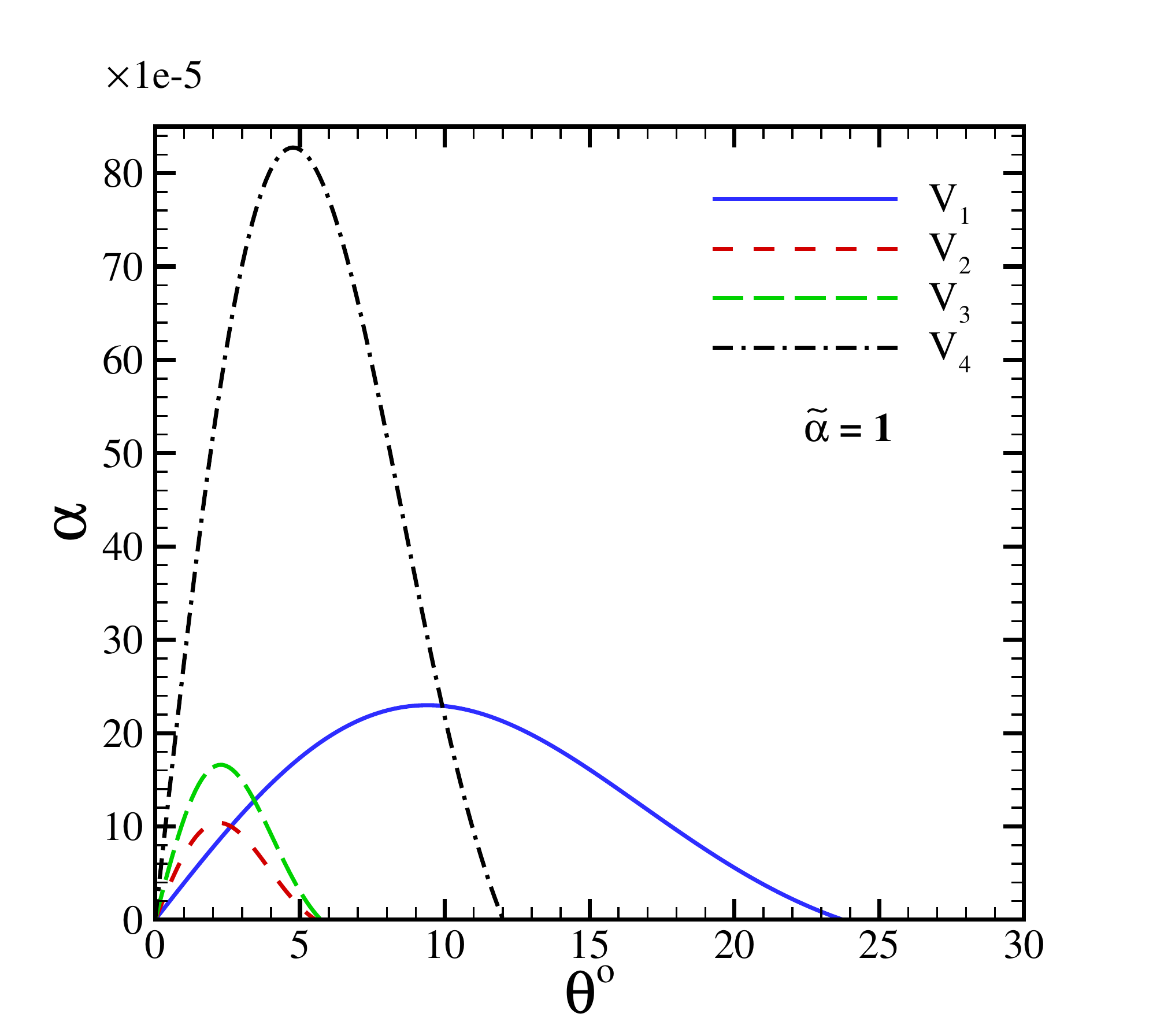}
\end{center}
\begin{center}
\caption{Similar to Figure~\ref{fig1} but for modified $\Lambda$LTB subvoids with  $\tilde\alpha=1$.}\label{fig11}
\end{center}
\end{figure}


The modified $\Lambda$LTB model is constructed by introducing a new parameter $\tilde\alpha$ in the potential profile, 
\begin{eqnarray}\label{eq:phi1}
\Phi(\tau,r)&=&\Phi_0\big(1-\tilde\alpha\frac{r^2}{r_0^2} \big)\exp\left[ -\frac{r^2}{R_{\rm V}^2}\right] \nonumber\\
&&\times {}_2F_1\left[1,\frac{1}{3},\frac{11}{6},\frac{-\Omega_{\Lambda}}{\Omega_{\bf m}(1+z)^3}\right]
\end{eqnarray}
that leads to the extended density profile 
\begin{eqnarray}
\delta(r,z_{\rm V})&=&\delta_{\rm V} g(z_{\rm V})  \exp\left(-\frac{r^2}{R_{\rm V}^2}\right) \nonumber \\
& &\times\left(1-\frac{2+7\tilde\alpha}{3+3\tilde\alpha}\frac{r^2}{R_{\rm V}^2}  +  \frac{2\tilde\alpha}{3+3\tilde\alpha}\frac{r^2}{R_{\rm V}^4} \right). \label{eq:delta_subv1}
\end{eqnarray}
The corresponding ISW and lensing potential imprints, valid for $0\le\tilde\alpha<2$,  are  calculated to be 
\begin{eqnarray} \label{eq:ltb1}
\delta T_{\rm rs}(\theta) &\approx&  \frac{3\sqrt{\pi}}{22} \delta_{\rm V}\frac{H(z_{\rm V})\Omega_\Lambda~ {}_2F_1\left[2,\frac{4}{3},\frac{17}{6},\frac{-\Omega_{\Lambda}}{\Omega_{\bf m}(1+z_{\bf V})^3}\right]}{H_0(1+z_{\rm V})^4~_2F_1\left[1,\frac{1}{3},\frac{11}{6},\frac{-\Omega_{\Lambda}}{\Omega_{\bf m}}\right]}  \nonumber \\ 
 && \times\left(H_0R_{\rm V}/c\right)^3 \left(\frac{2-\tilde\alpha}{2}-\tilde\alpha\frac{\theta^2}{\theta_{\rm V}^2}\right)   \exp\left(-\frac{ \theta^2}{\theta_{\rm V}^2} \right) \nonumber \\
&& \times\left(1+{\rm erf}\left[\frac{cz_{\rm V}}{H(z_{\rm V})R_{\rm V}} \right]\right) 
\end{eqnarray}
\citep{fin14} and 
\begin{eqnarray}
\alpha(\theta)&\approx&-\delta_{\rm V}\frac{D_{\rm LS}D_{\rm L}}{D_{\rm S}}\frac{2R_{\rm V}H_0^2\Omega_m}{c^2}\frac{{}_2F_1\left[1,\frac{1}{3},\frac{11}{6},\frac{-\Omega_{\Lambda}}{\Omega_{\bf m}(1+z_{\rm V})^3}\right]}{{}_2F_1\left[1,\frac{1}{3},\frac{11}{6},\frac{-\Omega_{\Lambda}}{\Omega_{\bf m}}\right]}  \nonumber \\ 
&&\times  {\left[ \tilde{\alpha} +\left(1-\tilde{\alpha}\frac{\theta^2}{\theta_{\rm V}^2}\right)\right]\theta \exp\left(-\frac{\theta^2}{\theta_{\rm V}^2}\right) }, \label{eq:alfa:ltbmv}
\end{eqnarray}
respectively.
Obviously, $\tilde\alpha=0$ reproduces the basic $\Lambda$LTB profiles.
For the modified templates we take $\tilde\alpha=1$.
 As the void parameters, we use the estimations of Table~\ref{table:subv}. It should be noted that these parameters are specifically fitted for the basic $\Lambda$LTB density profiles. However, the change in $\tilde\alpha$ does not drastically alter the overall density profiles, and therefor we assume the same set of parameters could be used as a first acceptable approximation to the modified $\Lambda$LTB model.  
Figure~\ref{fig11} illustrates the induced radial profiles of the  imprints. 
It is interesting to note the sensitivity of the induced anisotropies to relatively small changes in the density profiles. 
The suggested profiles are  compensated for all values of $\tilde\alpha$, with different $\tilde\alpha$s setting how quickly this compensation happens. 
We therefore follow a parallel approach by exploring sharply compensated subvoids, called the top-hat model, as an extreme scenario. These models have the potential of hot ring production and are introduced in the next section.
\subsubsection{Top-hat Subvoids}\label{sec:martinez}
\cite{ino06,ino07} proposed that the observed CMB CS can be produced by  a single huge  compensated void. In particular, they showed that a sharply compensated spherically symmetric void (i.e., one surrounded by a thin shell of matter that contains all the matter supposed to be in the void) can explain the CS in the microwave sky. As already discussed, their suggested void parameters were later ruled out by observations.
Here we investigate the possibility that multiple shallow subvoids with this assumed template placed at lower redshifts could source the observed CS anomaly. 

\cite{martinez1990anisotropies} and \cite{mar90} used approximations to  Einstein field equations for the linear non-static potential generated by nonlinear density fluctuations to study the  propagation of light and derive relatively simple expressions for the induced anisotropies in the CMB sky. Assuming a compensated spherical  void as the structure with density contrast $\delta_{\rm V}$ placed at redshift $z_{\rm V}$, with a length scale much smaller than the Hubble radius $D_H$, they found the following approximation for the generated anisotropy  
by the gravitational redshift of photons\footnote{The result in \cite{mar90}  is for an Einstein-de Sitter Universe. We follow \cite{cru08} to extend to the $\Lambda$CDM Universe with straightforward replacements.}, 
\begin{eqnarray}\label{eq:rsV}
 \delta T_{\rm rs}(\theta) &\approx&  - \delta_{\rm V}  \frac{16}{3}\left(\frac{\theta_{\rm V}}{1-d\cos \theta}\right)^3\nonumber\\
&& \sqrt{1-\frac{d^2\sin^2 \theta}{\theta_{\rm V}^2}}
\left(\frac{9}{2}\gamma -4 +\frac{d^2\sin^2 \theta}{\theta_{\rm V}^2}\right).\nonumber\\
\end{eqnarray}
  Here $\theta_{\rm V}  \propto t^\gamma$ is the angular size of the void corresponding to the comoving void radius $R_{\rm V}$
 where $\gamma\equiv E(z)\int_z^\infty (1+z')^{-1}E(z')^{-1} dz'$ describes  the propagation of the shell and $E(z)\equiv H(z)/H_0$.
The parameter $d$ represents the comoving distance of the observer to the center of the shell in units where $3t_0=1$ and $t_0$ is the age of the Universe. %
 Also, $\theta$ represents the angular distance between the direction of the observation and  the center of the structure. Note that Eq.~\ref{eq:rsV}  is valid only for $\theta\le \sin^{-1} \left( \theta_{\rm V}/d\right)$. Above this scale the net effect of the compensated void vanishes. 

\begin{figure}
\begin{center}
\includegraphics[scale=0.4]{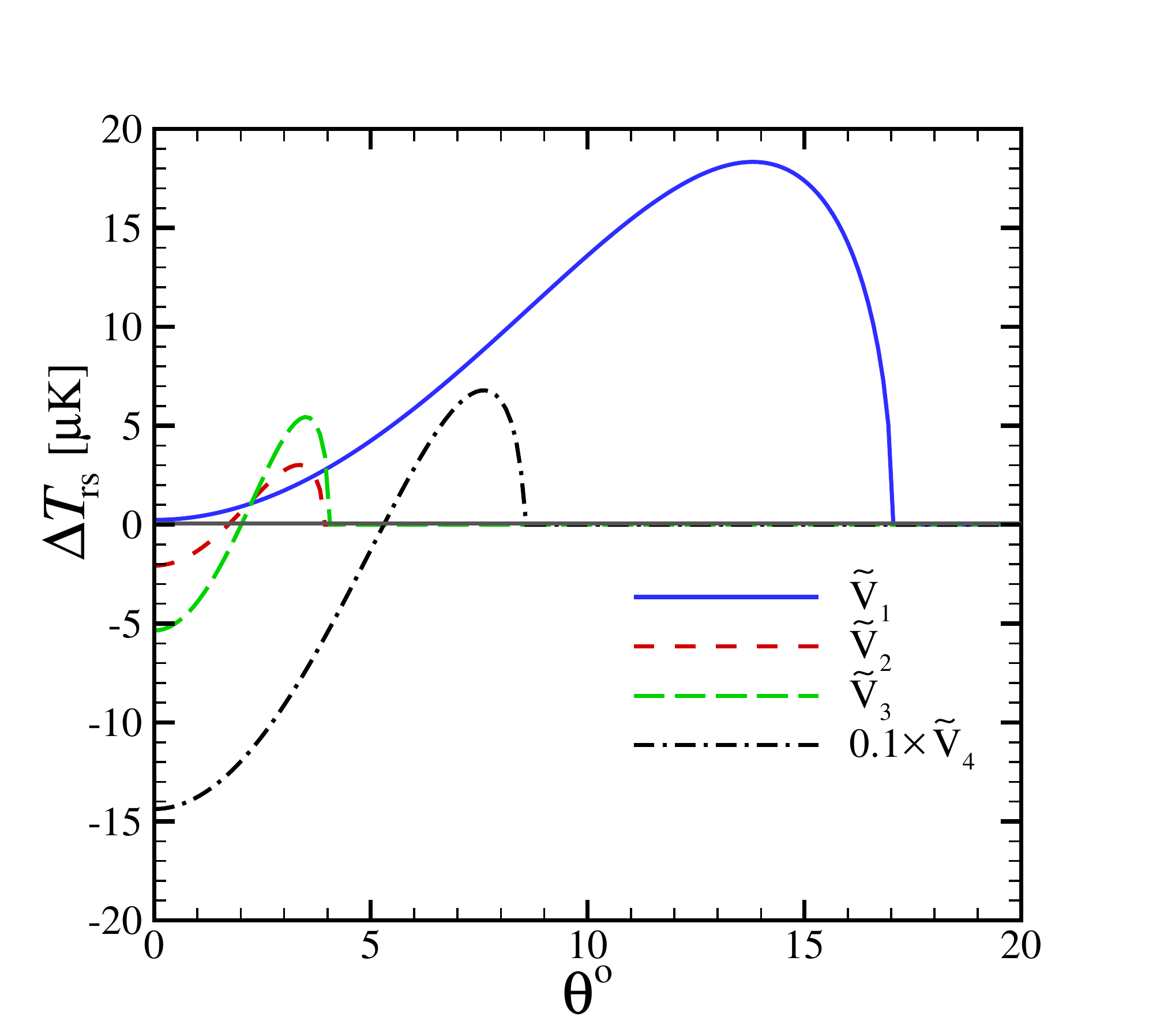}
\includegraphics[scale=0.4]{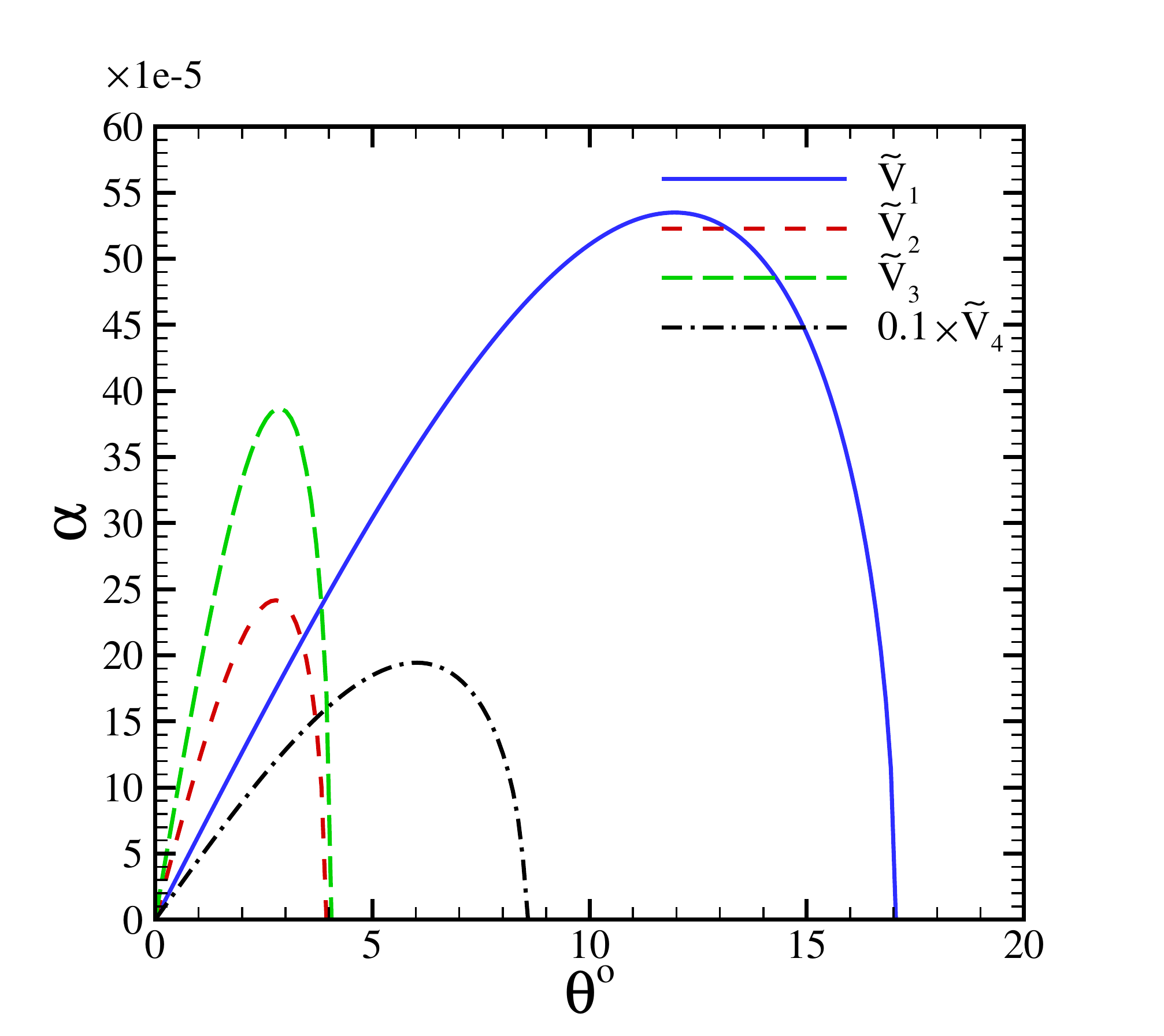}
\end{center}
\begin{center}
\caption{Similar to Figure~\ref{fig1} but for top-hat subvoids.}\label{fig2}
\end{center}
\end{figure}
 With the above assumptions for the void,  the deflection angle
 is shown to be well-approximated by \citep{mar90}
\begin{equation}\label{eq:lensV}
\alpha (\theta) \approx  - 8 \delta_{\rm V} \theta_{\rm V} d \sin \theta \sqrt{1-\frac{d^2\sin^2 \theta}{\theta_{\rm V}^{2}}}.
\end{equation}
In this work we assume the same characterizations for the top-hat subvoids as the $\Lambda$LTB ones, presented in Table~\ref{table:subv}.
It should be noted that these void specifications are derived through a fitting procedure of the galaxy distribution to the density profiles of the basic $\Lambda$LTB voids, modeled by Eq.~\ref{eq:delta_subv}. However, it is easy to see that 
the $\Lambda$LTB density profiles can be sharply approximated by top-hats in their underdense regions. We therefore do not expect significant shifts in the measurements of void specifications for the top-hat profiles compared to the $\Lambda$LTB ones.   
The radial profiles for the gravitational redshift of CMB photons induced by these structures are illustrated in the upper panel of Figure~\ref{fig3}, and their lensing deflection angles in the lower panel. 
It is interesting to  note that the compensated top-hat profiles can produce hot spots ($\tilde{\rm V}_1$) and rings ($\tilde{\rm V}_2$--$\tilde{\rm V}_4$), depending on the void redshift and size. See Eq.~\ref{eq:rsV}.
\subsection{Collapsing Cosmic Texture}\label{sec:texture}
Among the proposed explanations for the observed CMB CS is a collapsing cosmic texture \citep{cru07,cru08}. 
Cosmic textures are a type of topological defect possibly formed in phase transitions at early times, associated with symmetry breaking of certain models of high-energy physics.
As the Universe cools, the scalar fields present in the Higgs mechanism responsible for the symmetry breaking  acquire new non-zero expectation values and therefore form a vacuum manifold with a non-trivial topology. 
This can be achieved, e.g., when a global $\mathcal{O}(N)$ symmetry is broken spontaneously by $N$ scalar fields. 
In this case, for $N=4$ the corresponding vacuum manifold is a 3-sphere and the defects are called textures \citep{muk05}.   

\begin{figure}
\begin{center}
\includegraphics[scale=0.4]{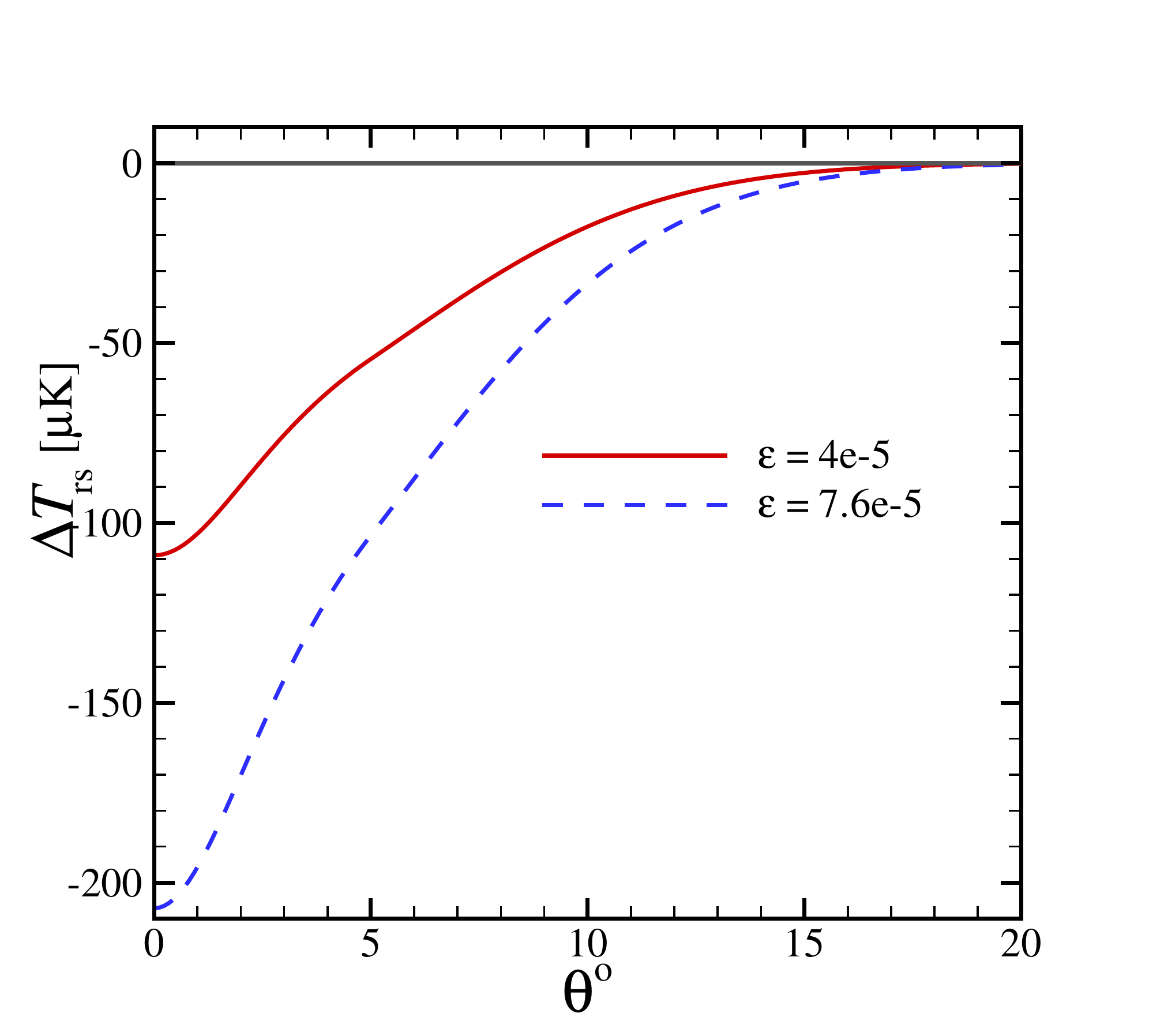}
\includegraphics[scale=0.4]{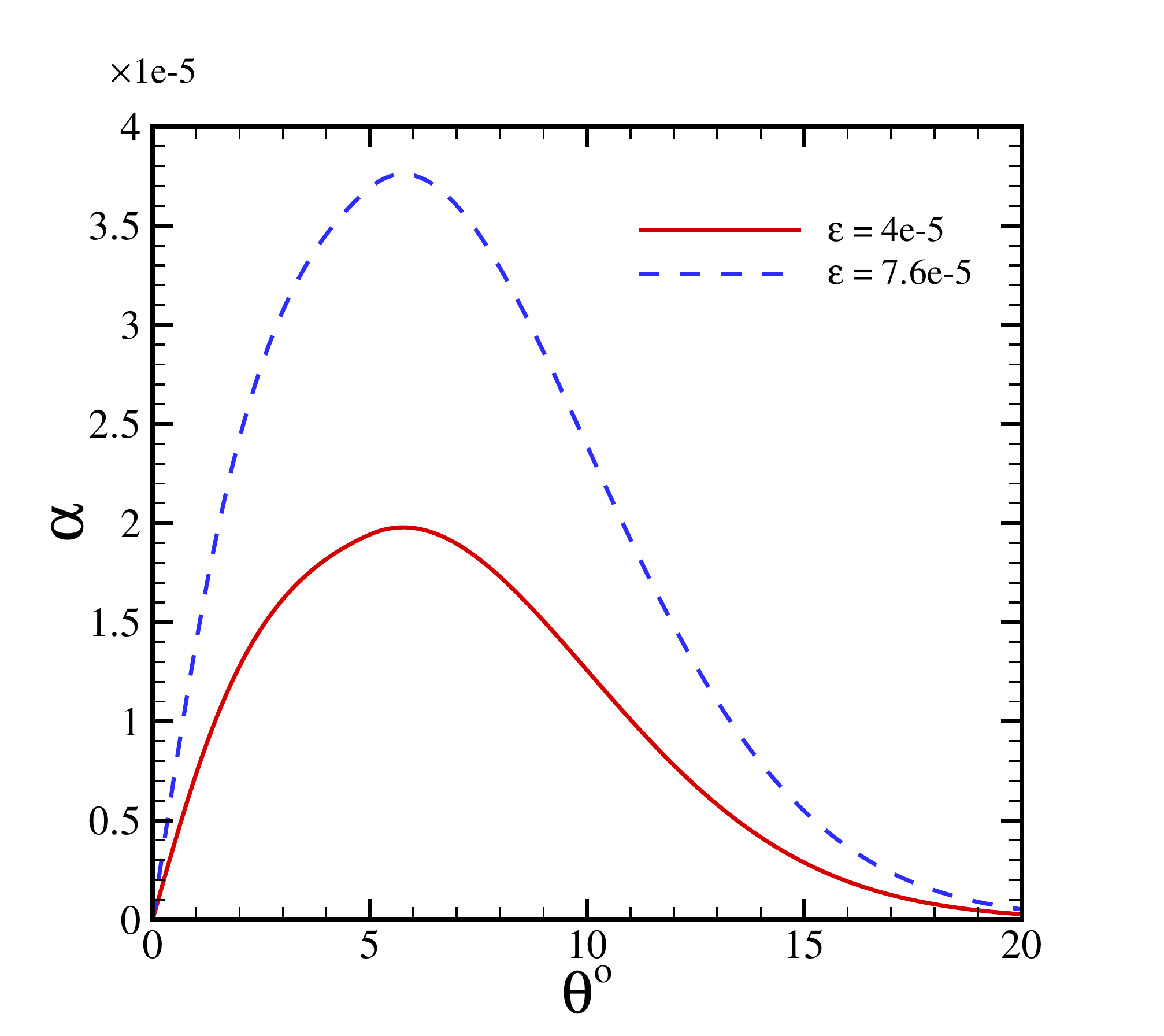}
\end{center}
\caption{Up: The radial profiles of temperature anisotropies due to gravitational redshift of a collapsing cosmic texture. Bottom: The corresponding radial profiles of the lensing deflection angle. }\label{fig3}
\end{figure}

Textures are unstable and after formation, textures first undergo collapse due to causal contact with their surrounding, and subsequently unwind when the gradient of the field energy surpasses the energy necessary to restore the global symmetry and emit outgoing massless radiation \citep{tur89,tur09,vil00}.
The photons passing through the non-static gravitational potential of a collapsing (expanding) texture would experience gravitational redshift (blueshift) and therefore a decrease (an increase) in their temperature.
 Such temperature anisotropies are approximated by \citep{pen94}

\begin{equation} \label{eq:rsT}
 \delta T_{\rm rs}(\theta)= - \epsilon \frac{1}{\sqrt{1+4\frac{\theta^2}{\theta^2_{\rm T}}}}~~~ \theta \le \theta_{*}.
\end{equation}
 Here $\theta$ represents the angular separation of the direction of the observation from the CS center as before  and $\theta_*\equiv\sqrt{3}/2\theta_{\rm T}$ where $\theta_{\rm T}$ is the characteristic angular size of the texture.  
 We have used the minus sign on the right hand side since to have  a cold spot redshifted (and not blueshifted) photons are required.  
 The $\epsilon$ is associated with the energy scale, $\psi_0$,  of the symmetry breaking phase transition through $\epsilon=8\pi^2G\psi_0^2$.
The radius $\theta_{\rm T}$ is determined by the dynamics of the Universe, as well as the redshift of the texture $z_{\rm T}$, through 
\begin{equation} \label{eq:RT}
\theta_{\rm T}=\frac{2\sqrt{2}\kappa(1+z_{\rm T})}{E(z_{\rm T})\int_0^{z_{\rm T}}dz/E(z)},
\end{equation}  
  where $\kappa$ is a fraction of unity set from simulations.
 It should be noted that the approximate profile of Eq.~(\ref{eq:rsT})  is only valid for temperature anisotropies up to $\theta \approx \theta_{\rm T}$. It can be extended to larger separations assuming the continuity of the profile and its first derivative at $\theta_{\rm T}$. 
 In this work we follow \cite{vie10} and use the following smooth extension for the template,
\begin{equation} \label{eq:rsTe}
 \delta T_{\rm rs}(\theta)= - \frac{\epsilon}{2}  e^{-(\theta^2-\theta_{*}^2)/2\theta^2_{\rm T}}
 %
\end{equation}
In addition to generating anisotropies in the microwave sky by gravitational redshift, the texture potential also acts as a converging lens  and bends the trajectories of the photons passing through it. 
 For the collapsing cosmic texture, the deflection angle  is modeled  as \citep{dur92,das09}
\begin{equation}\label{eq:lensT}
\alpha(\theta)= A_{\rm T}\frac{\theta}{\sqrt{1+4\frac{\theta^2}{\theta^2_{\rm T}}}}.
\end{equation}
 Here $A_{\rm T}\equiv\frac{2\sqrt{2}\epsilon}{\theta_{\rm T}}\frac{D_{\rm LS}}{D_{\rm S}}$, with $D_{\rm S}$ and $D_{\rm LS}$ representing the comoving distances of the source  to the observer and to the lens, respectively.
 We  follow the template fitting of earlier works based on WMAP data \citep{cru07} to adopt $\epsilon=4\times 10^{-5}$ and $\theta_{\rm T}\approx 5^\circ$. This places the texture at redshift $z_{\rm T}\sim 6$ and gives
$A_{\rm T}\approx 4.45 \times 10^{-4}$.
The upper panel of  Figure~\ref{fig3} represents the radial profile of the redshift-induced anisotropies, while the lower panel shows the  deflection angle due to the collapsing cosmic texture discussed above.
 \section{Analysis}\label{anal}
In the previous section we discussed the theoretical framework for the contribution to CMB anisotropies from the gravitational effects of multiple subvoids and a collapsing cosmic texture. We now explain our analysis method  with the aim to investigate the detectability (and consistency) of the traces  of these structures by the  {\it Planck} temperature data.
The datasets  consist of  discs  with radii $R=10^{\circ}$ and $R=20^{\circ}$, centered at the center of the CS, cut from the {\it Planck} \texttt{SMICA}  temperature map\footnote{https://pla.esac.esa.int/\#home} \citep{planckcomponent}.
The results presented here are with the HEALPix\footnote{https://healpix.sourceforge.io} resolution characterized by $N_{\rm side}=256$ for $R=10^{\circ}$ and $N_{\rm side}=128$ for $R=20^{\circ}$, corresponding to  pixel sizes of $\approx 14'$ and $\approx 28'$ respectively.  

We first consider the case for a single structure (either the sum of the imprints of multiple subvoids or a cosmic texture).
For a joint analysis of its gravitational redshift and lensing traces, assume that the observed temperature anisotropy of the patch can be modeled as
\begin{equation}
\delta T_{\rm obs}=\delta T_{\rm pri}+\mathcal{A}_{\rm rs}\delta T_{\rm rs} + \mathcal{A}_{\rm ln} \delta T_{\rm ln} +n 
\label{eq:dT}
 \end{equation}
where $\delta T_{\rm obs}$ and $\delta T_{\rm pri}$ represent the observed and  primordial CMB anisotropies, respectively. The contribution from the instrumental noise is represented by $n$. 
The template $\delta T_{\rm rs}$ denotes the template of the redshifted signal  produced by some decaying potential, here due to an expanding void or a collapsing texture.
The $\delta T_{\rm ln}$  describes the fluctuations due to the lensing of  CMB photons by the same source  (see Eq.~\ref{eq:lens}) characterized by the deflection angle ${\mathbf \alpha}(\theta)$ modeled by Eq.~\ref{eq:alfa:ltbmv}, ~\ref{eq:lensV} or ~\ref{eq:lensT}.
The corresponding amplitudes for the templates, $\mathcal{A}_{\rm rs}$ and $\mathcal{A}_{\rm ln}$, are free parameters to be simultaneously estimated from the data, and are expected to agree  within the error bars if our assumed scenario of the source of the CS makes a consistent picture.
Disagreements would then hint to possible inconsistencies in the ISW and lensing templates, improper parameter values  fixed {\it a priori} for template construction, or may even challenge the plausibility of the model. 
Within this framework, the likelihood of data $\mathcal{L}$, given the CS parameter pair $(\mathcal{A}_{\rm ln}, \mathcal{A}_{\rm rs})$, can be expressed through
\begin{equation}\label{eq:like}
-2\ln\mathcal{L}(g|\mathcal{A}_{\rm ln}, \mathcal{A}_{\rm rs})=g ^T {\bf C}^{-1}g + {\rm constant}
\end{equation}
where $g\equiv g(\mathcal{A}_{\rm ln}, \mathcal{A}_{\rm rs})=\delta T_{\rm obs}-\left[\mathcal{A}_{\rm rs}\delta T_{\rm rs} +\mathcal{A}_{\rm ln} \delta T_{\rm ln}\right]$ is the  Gaussian signal left when the CS contribution is properly subtracted from the observed data. 
We have made the simplifying assumption that the background primordial signal as well as the instrumental noise can be considered Gaussian. 
We construct the theoretically expected pixel-pixel covariance matrix for this Gaussian part, ${\bf C}_{pp'}\equiv \langle g_p g^T_{p'}\rangle$, from the full sky (except for the mask) {\it Planck}  temperature map. More specifically, we build the correlation function $C(\rho)$ where $\rho$ is the angular separation of any  pixel pair on the sphere, and use $C(\rho)$ to construct the ${\bf C}$ matrix for the desired patch. 
The best-fit values of the parameter pair ($\mathcal{A}_{\rm rs}^{\rm bf}, \mathcal{A}_{\rm ln}^{\rm bf}$) maximize their posterior probability distribution. 
Assuming uniform prior on the parameters in the vicinity of this best-fit point, the pair ($\mathcal{A}_{\rm rs}^{\rm bf}, \mathcal{A}_{\rm ln}^{\rm bf}$)  would also maximize the likelihood
\begin{equation}
\frac{\partial \mathcal{L}}{\partial \mathcal{A}_{\rm rs}}\bigg|_{\rm bf}=0 ~,~ \frac{\partial \mathcal{L}}{\partial \mathcal{A}_{\rm ln}}\bigg|_{\rm bf}=0
\end{equation}
yielding
\begin{eqnarray}
\mathcal{A}_{\rm rs}^{\rm bf}=\beta [(\delta T_{\rm ln}^\dagger {\bf C}^{-1} \delta T_{\rm ln}) (\delta T_{\rm rs}^\dagger {\bf C}^{\rm -1} \delta T_{\rm obs})\nonumber\\
-(\delta T_{\rm rs}^\dagger {\bf C}^{\rm -1} \delta T_{\rm ln})(\delta T_{\rm ln}^\dagger {\bf C}^{\rm -1} \delta T_{\rm obs})] \nonumber \\
\mathcal{A}_{\rm ln}^{\rm bf}=\beta[ (\delta T_{\rm rs}^\dagger {\bf C}^{-1} \delta T_{\rm rs})(\delta T_{\rm ln}^\dagger {\bf C}^{\rm -1} \delta T_{\rm obs})\nonumber\\
-(\delta T_{\rm ln}^\dagger {\bf C}^{\rm -1} \delta T_{\rm rs})(\delta_{\rm rs}^\dagger {\bf C}^{\rm -1} \delta T_{\rm obs})] \label{eq:Asing}
\end{eqnarray}
where
\begin{eqnarray}
\beta^{-1}\equiv(\delta T_{\rm rs}^\dagger {\bf C}^{-1} \delta T_{\rm rs})(\delta T_{\rm ln}^\dagger {\bf C}^{\rm -1} \delta T_{\rm ln})\\ \nonumber
-(\delta T_{\rm rs}^\dagger {\bf C}^{\rm -1} \delta T_{\rm ln})(\delta T_{\rm ln}^\dagger {\bf C}^{\rm -1} \delta_{\rm rs}).
\end{eqnarray}
One could  use the width of the likelihood surface, i.e., the square root of the diagonals of the Fisher inverse, to get an estimate of the  measurement error,
\begin{equation}
\sigma^2_{\rm rs}=\frac{\partial^2 \mathcal{L}}{\partial \mathcal{A}^2_{\rm rs}}\bigg|_{\rm bf}~~,~~\sigma^2_{\rm ln}=\frac{\partial^2 \mathcal{L}}{\partial A_{\rm ln}^2}\bigg|_{\rm bf}.
\end{equation}
They yield reliable estimates of the uncertainties for close-to-Gaussian distributions, and  a lower limit otherwise. 
Using Eq.~(\ref{eq:like}) we get
\begin{eqnarray} \label{eq:sig}
\sigma_{\rm rs}&=&\sqrt{(\delta T_{\rm ln}^\dagger {\bf C}^{-1} \delta T_{\rm ln})/\beta},\nonumber\\
\sigma_{\rm ln}&=&\sqrt{(\delta T_{\rm rs}^\dagger {\bf C}^{-1} \delta T_{\rm rs})/\beta}.
\end{eqnarray}

In the next section we also consider cases where the amplitudes of the individual subvoids are considered as free parameters and their consistency with the suggested values of Table~\ref{table:subv} is investigated. 
The extension of the above framework to these cases with several varying amplitudes is straightforward,
\begin{equation}
\delta T_{\rm obs}=\delta T_{\rm pri}+\vec{\mathcal{A}}_{\rm rs}\cdot \vec{\delta T_{\rm rs}} + \vec{\mathcal{A}}_{\rm ln} \cdot \vec{\delta T_{\rm ln}} +n 
\label{eq:dT}
 \end{equation}
where $ \vec{\delta T_{\rm rs}}$ and $\vec{\delta T_{\rm ln}}$ are the  arrays of the gravitational redshift  and lensing templates and $\vec{\mathcal{A}}_{\rm rs}$ and $\vec{\mathcal{A}}_{\rm ln}$ are the arrays of their amplitudes. 
We see in the next section that  the correlation between lensing and redshifted signal is tiny. We therefore neglect it here. 
The best-fit amplitudes and their Fisher-based errors can be found similar to the previous case of a single structure, and we get
\begin{equation}
\vec{\mathcal{A}}_{\diamond}^{\rm bf}= M^{-1}_{\diamond}\cdot \vec{d}_{ \diamond}, ~~~ \vec{\sigma}_{\diamond}={\rm diag}(M^{-1}_{\diamond})
\label{eq:vecA}
 \end{equation}
 where 
 \begin{equation}
 \vec{d}_{\diamond}= \delta\vec{T}_{\diamond}  {\bf C}^{-1} \delta T_{\rm obs}, ~~~ M_{\diamond}=\delta\vec{T}_{\diamond} {\bf C}^{-1}\delta\vec{T}_{\diamond}  
 \end{equation}\label{eq:dMx}
and  the symbol ${\diamond}$ represents either rs or ln.
Eq.~\ref{eq:vecA} reduces to Eq.s~\ref{eq:Asing} and \ref{eq:sig} for the case of a texture or a single void.
  
\section{results}\label{res}
We follow the method  explained in the previous section to measure the amplitudes of the gravitational redshift and lensing templates for the candidates introduced in Section~\ref{candid}. 
We calculate the  Fisher-estimated errors for $\vec{\mathcal{A}}_{\rm rs}$ and $\vec{\mathcal{A}}_{\rm ln}$ with all other standard parameters fixed.
As stated in the introduction, an important feature of the observed cold spot in CMB maps is the hot ring, with $R\approx 15^\circ$, surrounding the inner cold region. 
Therefore, a realistic explanation of the CS requires capturing this ring as well. 
We explore how extending the patch radius to $R=20^\circ$ that well includes the ring would affect the results. For these larger discs, we reduce the resolution to $N_{\rm side}=128$ to reduce the computational cost. 
\begin{table}
\centering
\caption{The amplitudes of the gravitational redshift $\mathcal{A}_{\rm rs}$ and the lensing amplitude $\mathcal{A}_{\rm lens}$ for the sum of the multiple subvoids of Table~\ref{table:subv}, simultaneously measured by the {\it Planck} \texttt{SMICA} temperature map\citep{planckcomponent}. The three sets of subvoids, ${\rm V}(\tilde{\alpha}=0)$, ${\rm V}(\tilde{\alpha}=1)$ and $\tilde{\rm V}$ correspond to the models of Sections~\ref{sec:mackenzie} and ~\ref{sec:martinez}.}
\begin{tabular}{cccccc}
\hline\hline\\
  &&$\sum{\rm V}_i (\tilde\alpha=0) $  & $\sum{\rm V}_{i}(\tilde\alpha=1$) &$\sum\tilde{\rm V}_i $ &Texture \\ \hline
         	\multirow{2}{*}{$\mathcal{A}_{\rm rs}$}& $  R=10^\circ$ & $3.4 \pm 1.5 $   & $16.4 \pm 3.9 $	& $ 0.1 \pm 0.1$ & $1.5 \pm 0.5 $ \\	
		 &$R=20^\circ $ &$ 5.4 \pm 1.4  $  & $14.4 \pm3.8 $ &$ 0.3 \pm 0.1 $ & $1.9 \pm 0.5$ \\ \hline 
         	\multirow{2}{*}{ $\mathcal{A}_{\rm ln}$}& $  R=10^\circ$ &$<0.1  $ & $<0.1$ &$ <0.02 $ & $< 1.7$ \\	
		 &$R=20^\circ $ &$<0.2 $ & $<0.1$ & $ <0.1 $  &  $< 6.2 $\\ \hline 	
       \end{tabular}\label{table:resv}
\end{table}

\begin{figure}
\begin{center}
\includegraphics[scale=0.4]{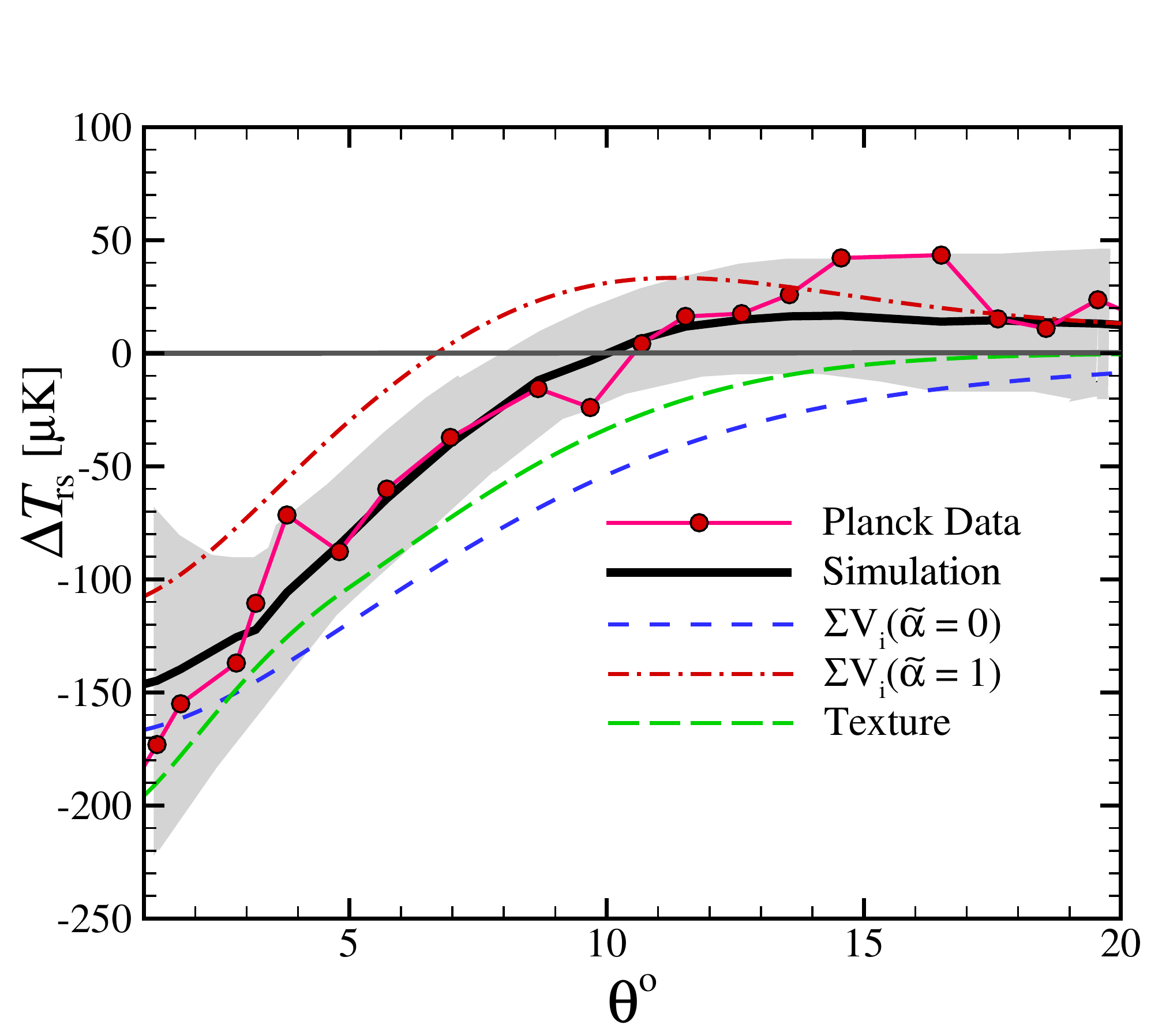}
\end{center}
\caption{The radial profiles of the estimated gravitational redshift imprints of the multiple subvoids and texture (see Table~\ref{table:resv}). The {\it Planck} data points, the simulated profile (the solid black curve) and their corresponding error bars are taken from Figure 30 of \cite{pl18stat}.}\label{fig4}
\end{figure}

Table~\ref{table:resv} presents the measured amplitudes of the sum of the subvoid templates of Figures~\ref{fig1}-\ref{fig2}, corresponding to $\sum {\rm V}_i(\tilde\alpha=0)$, $\sum {\rm V}_{i}(\tilde\alpha=1)$ and $\sum \tilde{\rm V}_i$ respectively, as well as the texture of Figure~\ref{fig3}, for patches with $R=10^\circ$ and $R=20^\circ$ centered at the CS.
The amplitudes for the gravitational redshift are measured to be non-zero with different levels of significance, depending on the patch size, for void  and the texture templates. 
The radial profiles of the predicted imprints for the different sets of candidates are shown in Figure~\ref{fig4} and compared with the measured {\it Planck} profile for patches with $R=20^{\circ}$.  It is worth to note that the model parameters, and the resulting best-fit profiles, are estimated by maximizing the likelihood of Eq.~\ref{eq:like} with the full pixel covariance matrix taken into account. 
For the $R=20^\circ$ patches, the negative correlations of large pixel-pair separations can impact the results in a non-trivial way.
We verified this effect by ignoring the correlations and limiting the likelihood to the diagonal terms, which is often used for a quick estimation of the amplitude, or in visual comparisons with the data profile. We noticed small changes in the measured amplitudes, depending on the patch size and template. We therefor emphasize that the measurements of Table~\ref{table:resv} are the reliable results with the full information of the CMB temperature anisotropies taken into account (under the assumption of Gaussian primordial fluctuations).

The measured amplitudes for voids are in tension, again with different levels of significance, with $\mathcal{A}_{\rm rs}=1$  which is the expected level if the assumed template is truly capable of describing the CS. 
The observed high amplitudes for $\sum{\rm V}_i(\tilde {\alpha}=0)$ and $\sum{\rm V}_{i}(\tilde {\alpha}=1)$ are to be expected, given the low values of anisotropies produced by these templates (see Figures~\ref{fig1} and~\ref{fig11} in the central region of the CS where an anisotropy of $\sim 100-150$ $\mu$K is required). For the $\sum\tilde{\rm V}_i$ amplitudes, however, simple visual inspection should be treated with care as the templates cross the $\delta T$ axis at small angles and assigning high amplitudes to the template would lead to large positive anisotropies at small angles, which is inconsistent with data.

It should be noted that the sum of the basic $\Lambda$LTB voids, unlike the modified templates with $\tilde\alpha=1$, suffers from being incapable of reproducing the observed hot ring as the individual templates all fade to zero at large radii. 
The sum of the top-hat voids, on the other hand,  is dominated by $\tilde{\rm V}_1$ on large radiaa (Figure~\ref{fig2}), with little contribution close to the CS 
center and positive contribution at larger radii. 
A blind and straightforward interpretation of the measured amplitudes and their deviation from unity is to assign  a different underdensity for the void obtained through multiplying the measured $\mathcal{A}_{\rm rs}$ by the fiducial values of Table~\ref{table:subv}, if it does not exceed $-1$. 
However, given the observational support for the fiducial underdensities, assigning a different $\delta_{\rm V}$ to the voids is hard to justify. See the discussion in Section~\ref{concl}.

The measured lensing amplitudes are consistent with zero for all of these templates, and in tension with the  redshifted signals of the voids. This inconsistency between the ISW and lensing amplitudes would challenge the above interpretation of the measured ISW amplitudes as a multiplier to the assumed underdensity.  We also, not surprisingly, find negligible correlation between the amplitudes of the gravitational redshift and lensing templates.  

The redshift imprint of texture translates to bounds on the energy scale of the corresponding phase transition through $\epsilon=(6.0\pm 2.0)\times 10^{-5}$ ($1\sigma$) 
in the inner cold region and 
to $\epsilon=(7.6\pm 2.0)\times 10^{-5}$ ($1\sigma$) 
when the outer ring is included. It should however be noted that the texture template smoothly goes to zero and, similar to the case of the basic $\Lambda$LTB subvoids, cannot reproduce the outer ring. Therefore, any conclusion about the consistency of the given texture profile with the observed CS should be taken with care. 

For comparison, we also consider a case where the subvoid amplitudes are all treated as free parameters and are simultaneously measured by data, together with the corresponding lensing amplitudes. 
This split analysis could be regarded as a consistency check of the model to make sure not only the sum, but also the individual void profiles, when superposed, are consistent with observations. 
We find all the ISW amplitudes to deviate significantly and at different levels from unity and all the lensing amplitudes are measured to be consistent with zero. 
The results of Table~\ref{table:resv} and these inconsistencies pose serious challenges in the interpretation of the subvoid templates as the origin of the CS in the $\Lambda$CDM framework.

\section{summary and discussion}\label{concl}
In this work we addressed the possibility of the CMB cold spot being
produced by the contributions of multiple subvoids or a collapsing cosmic texture. 
The subvoids correspond to the observations of  the 2dF-VST ATLAS Cold Spot galaxy redshift survey. 
The  goal was to
search for the signatures these candidates, as physical sources of the CS, would leave on CMB temperature anisotropies, as seen by $Planck$,  in the form of gravitational
redshift and lensing parameterized by  $\mathcal{A}_{\rm rs} $ and  $\mathcal{A}_{\rm ln}$ respectively. 
In the case of nonzero amplitudes for the templates, proper comparison of the amplitudes would shed light on the self-consistency of the model. 
One could even consider exploiting the information in the measured amplitudes to constrain various physical parameters characterizing the candidate templates. 

The non-zero amplitudes for the gravitational redshift signals of the multiple subvoids were found to deviate from unity. In the $\Lambda$LTB case, boosted anisotropies were required compared to the level expected from $\Lambda$CDM. The top-hat subvoids, on the other hand, required lower levels of anisotropies, due to being dominated by the large imprint of the forth subvoid.
The higher than expected amplitudes for the subvoid imprints were also previously claimed by other authors, e.g., see \cite{fin14} and \cite{mac17}.   
%
%
%
There was also no detection for the lensing signals for any of these structures 
and some of the upper bounds were in significant tension with the estimated redshifted signatures.

The conclusion for the texture is somewhat different. 
The measured amplitude corresponds to an energy scale of $(7.6\pm2.0)\times10^{-5}$ when both the inner cold and out hot regions are included (and lower when the hot ring is excluded). This is still below the current upper bounds on the energy scale of inflation set by the constraints on primordial gravitational waves and therefore consistent with the theoretical requirement of the symmetry breaking happening after the end of inflation \citep{fen12}.
%
 However, the observed hot ring cannot be explained by the texture profile.  Further conclusions about the viability of the texture hypothesis requires an explanation of this ring as well. The upper bound on the texture  lensing amplitude  is consistent with one, and therefore consistent with the gravitational redshift signal and the upper bound set by inflation. Therefore, higher resolution analysis is expected to independently shed more light on the texture assumption. 

%
%
%

We conclude that in the light of the $Planck$ data and in the standard cosmological framework of $\Lambda$CDM, the various assumed subvoid templates of this work cannot  reproduce the observed CS and the surrounding hot ring. 
The deviations of the amplitudes from unity,  the tension between the gravitational redshift and lensing amplitudes, and the inconsistency of the individual void amplitudes when allowed to freely vary with the amplitude of the sum of the subvoids are serious challenges in the interpretation of the observed subvoids as the main source of the CS in the $\Lambda$CDM framework and call for further theoretical considerations.


 
One could think of modifications to the models, or exploration in a broader parameter space in a limited region allowed by the current observational bounds set by fitting the void templates with galaxy surveys. 
Variations in the void shape can have significant impacts on the induced temperature profiles. For example, although the underdense region of the void templates used in this work are close, the difference in the compensating overdensities that surround the inner part lead to significantly different temperature profiles. 
Alternative approaches include hints to primordial origins for the CS \cite[e.g.,][]{ringeval2016large} or call for modification in the underlying cosmological model of $\Lambda$CDM. 
In particular, the size of the cold region in the top-hat voids is set by the angle where the last parenthesis in Eq.~\ref{eq:rsV} vanishes. This in turn is sensitive to the cosmology-dependent parameter $\gamma$. It is therefore expected that variations in the model of dark energy change the induced cold and hot profiles and their crossover and the resulting observational fit.  
%
If the observed  2dF-VST ATLAS subvoids are to explain the CMB CS in a different cosmological scenario,  not only the total ISW and lensing amplitudes, but also the simultaneously measured amplitudes of the individual subvoids should be consistent with one. 

Further observational probes such as imprints on the 21 cm signal  by the gravitational redshift and lensing of these structures 
could be used to independently assess the viability of these candidates as sources of the cold spot.


{\bf Acknowledgement} The numerical calculations of this work were carried out on the computing
cluster of the Canadian Institute for Theoretical Astrophysics (CITA), University of Toronto. The authors are also grateful to Alireza Vafaei Sadr for the help in preparing the {\it Planck} simulations required at some initial stages of this work.

\bibliography{cs}

\begin{thebibliography}{69}
\expandafter\ifx\csname natexlab\endcsname\relax\def\natexlab#1{#1}\fi

\bibitem[{Acquaviva \& Baccigalupi(2006)}]{acquaviva2006dark}
Acquaviva, V., \& Baccigalupi, C. 2006, Physical Review D, 74, 103510

\bibitem[{Ade {et~al.}(2016)Ade, Aghanim, Arnaud, Ashdown, Aumont, Baccigalupi,
  Banday, Barreiro, Bartolo, Basak, {et~al.}}]{ade2016planck}
Ade, P., {et~al.} 2016, Astronomy \& Astrophysics, 594, A21

\bibitem[{Ade {et~al.}(2014)}]{pl13}
Ade, P. A.~R., {et~al.} 2014, Astron. Astrophys., 571, A23

\bibitem[{Afshordi(2004)}]{afshordi2004integrated}
Afshordi, N. 2004, Physical Review D, 70, 083536

\bibitem[{Amendola {et~al.}(2008)Amendola, Kunz, \&
  Sapone}]{amendola2008measuring}
Amendola, L., Kunz, M., \& Sapone, D. 2008, Journal of Cosmology and
  Astroparticle Physics, 2008, 013

\bibitem[{Beck {et~al.}(2018)Beck, Csabai, Rácz, \& Szapudi}]{bec18}
Beck, R., Csabai, I., Rácz, G., \& Szapudi, I. 2018, Mon. Not. Roy. Astron.
  Soc., 479, 3582

\bibitem[{Bremer {et~al.}(2010)Bremer, Silk, Davies, \& Lehnert}]{bre10}
Bremer, M.~N., Silk, J., Davies, L. J.~M., \& Lehnert, M.~D. 2010, Mon. Not.
  Roy. Astron. Soc., 404, 69

\bibitem[{Cai {et~al.}(2010)Cai, Cole, Jenkins, \& Frenk}]{cai10}
Cai, Y.-C., Cole, S., Jenkins, A., \& Frenk, C.~S. 2010, Monthly Notices of the
  Royal Astronomical Society, 407, 201

\bibitem[{Carbone {et~al.}(2013)Carbone, Baldi, Pettorino, \&
  Baccigalupi}]{carbone2013maps}
Carbone, C., Baldi, M., Pettorino, V., \& Baccigalupi, C. 2013, Journal of
  Cosmology and Astroparticle Physics, 2013, 004

\bibitem[{{Colberg} {et~al.}(2005){Colberg}, {Sheth}, {Diaferio}, {Gao}, \&
  {Yoshida}}]{col05}
{Colberg}, J.~M., {Sheth}, R.~K., {Diaferio}, A., {Gao}, L., \& {Yoshida}, N.
  2005, \mnras, 360, 216

\bibitem[{Courtois {et~al.}(2017)Courtois, Tully, Hoffman, Pomarede, Graziani,
  \& Dupuy}]{cou17}
Courtois, H.~M., Tully, R.~B., Hoffman, Y., Pomarede, D., Graziani, R., \&
  Dupuy, A. 2017, Astrophys. J., 847, L6

\bibitem[{Cruz {et~al.}(2007)Cruz, Cayon, {Mart{\'{\i}}nez-Gonz{\'a}lez},
  Vielva, \& Jin}]{cru07a}
Cruz, M., Cayon, L., {Mart{\'{\i}}nez-Gonz{\'a}lez}, E., Vielva, P., \& Jin, J.
  2007, The Astrophysical Journal, 655, 11

\bibitem[{{Cruz} {et~al.}(2005){Cruz}, {Mart{\'{\i}}nez-Gonz{\'a}lez},
  {Vielva}, \& {Cay{\'o}n}}]{cru05}
{Cruz}, M., {Mart{\'{\i}}nez-Gonz{\'a}lez}, E., {Vielva}, P., \& {Cay{\'o}n},
  L. 2005, \mnras, 356, 29

\bibitem[{Cruz {et~al.}(2005)Cruz, {Mart{\'{\i}}nez-Gonz{\'a}lez}, Vielva, \&
  Cayon}]{cru04}
Cruz, M., {Mart{\'{\i}}nez-Gonz{\'a}lez}, E., Vielva, P., \& Cayon, L. 2005,
  Mon. Not. Roy. Astron. Soc., 356, 29

\bibitem[{{Cruz} {et~al.}(2008){Cruz}, {Mart{\'{\i}}nez-Gonz{\'a}lez},
  {Vielva}, {Diego}, {Hobson}, \& {Turok}}]{cru08}
{Cruz}, M., {Mart{\'{\i}}nez-Gonz{\'a}lez}, E., {Vielva}, P., {Diego}, J.~M.,
  {Hobson}, M., \& {Turok}, N. 2008, \mnras, 390, 913

\bibitem[{{Cruz} {et~al.}(2006){Cruz}, {Tucci}, {Mart{\'{\i}}nez-Gonz{\'a}lez},
  \& {Vielva}}]{cru06}
{Cruz}, M., {Tucci}, M., {Mart{\'{\i}}nez-Gonz{\'a}lez}, E., \& {Vielva}, P.
  2006, \mnras, 369, 57

\bibitem[{Cruz {et~al.}(2006)Cruz, Tucci, {Mart{\'{\i}}nez-Gonz{\'a}lez}, \&
  Vielva}]{cru06a}
Cruz, M., Tucci, M., {Mart{\'{\i}}nez-Gonz{\'a}lez}, E., \& Vielva, P. 2006,
  Mon. Not. Roy. Astron. Soc., 369, 57

\bibitem[{{Cruz} {et~al.}(2007){Cruz}, {Turok}, {Vielva},
  {Mart{\'{\i}}nez-Gonz{\'a}lez}, \& {Hobson}}]{cru07}
{Cruz}, M., {Turok}, N., {Vielva}, P., {Mart{\'{\i}}nez-Gonz{\'a}lez}, E., \&
  {Hobson}, M. 2007, Science, 318, 1612

\bibitem[{{Das} \& {Spergel}(2009)}]{das09}
{Das}, S., \& {Spergel}, D.~N. 2009, \prd, 79, 043007

\bibitem[{{Durrer} {et~al.}(1992){Durrer}, {Heusler}, {Jetzer}, \&
  {Straumann}}]{dur92}
{Durrer}, R., {Heusler}, M., {Jetzer}, P., \& {Straumann}, N. 1992, Nuclear
  Physics B, 368, 527

\bibitem[{Feeney {et~al.}(2012)Feeney, Johnson, Mortlock, \& Peiris}]{fen12}
Feeney, S.~M., Johnson, M.~C., Mortlock, D.~J., \& Peiris, H.~V. 2012, Physical
  Review Letters, 108

\bibitem[{Finelli {et~al.}(2016)Finelli, García-Bellido, Kovács, Paci, \&
  Szapudi}]{fin14}
Finelli, F., García-Bellido, J., Kovács, A., Paci, F., \& Szapudi, I. 2016,
  Mon. Not. Roy. Astron. Soc., 455, 1246

\bibitem[{Flender {et~al.}(2013)Flender, Hotchkiss, \& Nadathur}]{fle13}
Flender, S., Hotchkiss, S., \& Nadathur, S. 2013, Journal of Cosmology and
  Astroparticle Physics, 2013, 013

\bibitem[{Garcia-Bellido \& Haugboelle(2008)}]{gar08}
Garcia-Bellido, J., \& Haugboelle, T. 2008, Journal of Cosmology and
  Astroparticle Physics, 2008, 003

\bibitem[{García-Bellido {et~al.}(2011)García-Bellido, Durrer, Fenu, Figueroa,
  \& Kunz}]{gar11}
García-Bellido, J., Durrer, R., Fenu, E., Figueroa, D.~G., \& Kunz, M. 2011,
  Physics Letters B, 695, 26

\bibitem[{Granett {et~al.}(2010)Granett, Szapudi, \& Neyrinck}]{gra09}
Granett, B.~R., Szapudi, I., \& Neyrinck, M.~C. 2010, Astrophys. J., 714, 825

\bibitem[{Hotchkiss {et~al.}(2014)Hotchkiss, Nadathur, Gottlober, Iliev, Knebe,
  Watson, \& Yepes}]{hot14}
Hotchkiss, S., Nadathur, S., Gottlober, S., Iliev, I.~T., Knebe, A., Watson,
  W.~A., \& Yepes, G. 2014, Monthly Notices of the Royal Astronomical Society,
  446, 1321

\bibitem[{{Hoyle} \& {Vogeley}(2004)}]{hoy04}
{Hoyle}, F., \& {Vogeley}, M.~S. 2004, \apj, 607, 751

\bibitem[{Hu \& White(1996)}]{hu1996acoustic}
Hu, W., \& White, M. 1996, The Astrophysical Journal, 471, 30

\bibitem[{{Inoue} \& {Silk}(2006)}]{ino06}
{Inoue}, K.~T., \& {Silk}, J. 2006, \apj, 648, 23

\bibitem[{{Inoue} \& {Silk}(2007)}]{ino07}
---. 2007, \apj, 664, 650

\bibitem[{Kovetz \& Kamionkowski(2013)}]{kov13}
Kovetz, E.~D., \& Kamionkowski, M. 2013, Physical Review Letters, 110

\bibitem[{Kovács(2018)}]{kov17}
Kovács, A. 2018, Mon. Not. Roy. Astron. Soc., 475, 1777

\bibitem[{Kovács \& García-Bellido(2016)}]{kov15}
Kovács, A., \& García-Bellido, J. 2016, Mon. Not. Roy. Astron. Soc., 462, 1882

\bibitem[{Kovács \& Granett(2015)}]{kovb15}
Kovács, A., \& Granett, B.~R. 2015, Monthly Notices of the Royal Astronomical
  Society, 452, 1295

\bibitem[{Larson \& Wandelt(2004)}]{larson2004hot}
Larson, D.~L., \& Wandelt, B.~D. 2004, The Astrophysical Journal Letters, 613,
  L85

\bibitem[{Liddle \& Lyth(2000)}]{lid00}
Liddle, A.~R., \& Lyth, D.~H. 2000, Cosmological inflation and large-scale
  structure (Cambridge university press)

\bibitem[{{Mackenzie} {et~al.}(2017){Mackenzie}, {Shanks}, {Bremer}, {Cai},
  {Gunawardhana}, {Kov{\'a}cs}, {Norberg}, \& {Szapudi}}]{mac17}
{Mackenzie}, R., {Shanks}, T., {Bremer}, M.~N., {Cai}, Y.-C., {Gunawardhana},
  M.~L.~P., {Kov{\'a}cs}, A., {Norberg}, P., \& {Szapudi}, I. 2017, \mnras,
  470, 2328

\bibitem[{Marcos-Caballero {et~al.}(2016)Marcos-Caballero, Fernández-Cobos,
  Martínez-González, \& Vielva}]{mar16}
Marcos-Caballero, A., Fernández-Cobos, R., Martínez-González, E., \& Vielva, P.
  2016, Monthly Notices of the Royal Astronomical Society: Letters, 460, L15

\bibitem[{{Mart{\'{\i}}nez-Gonz{\'a}lez}
  {et~al.}(1990){Mart{\'{\i}}nez-Gonz{\'a}lez}, Sanz, \&
  Silk}]{martinez1990anisotropies}
{Mart{\'{\i}}nez-Gonz{\'a}lez}, E., Sanz, J., \& Silk, J. 1990, The
  Astrophysical Journal, 355, L5

\bibitem[{{{Mart{\'{\i}}nez-Gonz{\'a}lez}} \& {Sanz}(1990)}]{mar90}
{{Mart{\'{\i}}nez-Gonz{\'a}lez}}, E., \& {Sanz}, J.~L. 1990, \mnras, 247, 473

\bibitem[{Mostaghel {et~al.}(2018)Mostaghel, Moshafi, \&
  Movahed}]{mostaghel2018integrated}
Mostaghel, B., Moshafi, H., \& Movahed, S. 2018, Monthly Notices of the Royal
  Astronomical Society, 481, 1799

\bibitem[{Mukhanov(2005)}]{muk05}
Mukhanov, V. 2005, {Physical Foundations of Cosmology} (Oxford: Cambridge
  University Press)

\bibitem[{Nadathur {et~al.}(2014)Nadathur, Lavinto, Hotchkiss, \&
  Räsänen}]{nad14}
Nadathur, S., Lavinto, M., Hotchkiss, S., \& Räsänen, S. 2014, Phys. Rev., D90,
  103510

\bibitem[{Papai \& Szapudi(2010)}]{pap10a}
Papai, P., \& Szapudi, I. 2010, Astrophys. J., 725, 2078

\bibitem[{Papai {et~al.}(2011)Papai, Szapudi, \& Granett}]{pap10b}
Papai, P., Szapudi, I., \& Granett, B.~R. 2011, Astrophys. J., 732, 27

\bibitem[{{Patiri} {et~al.}(2006){Patiri}, {Betancort-Rijo}, {Prada}, {Klypin},
  \& {Gottl{\"o}ber}}]{pat06}
{Patiri}, S.~G., {Betancort-Rijo}, J.~E., {Prada}, F., {Klypin}, A., \&
  {Gottl{\"o}ber}, S. 2006, \mnras, 369, 335

\bibitem[{Peebles(1984)}]{peebles1984tests}
Peebles, P. 1984, ApJ, 284, 439

\bibitem[{Peiris(2014)}]{pei14}
Peiris, H.~V. 2014, Proceedings of the International Astronomical Union, 10,
  124

\bibitem[{{Pen} {et~al.}(1994){Pen}, {Spergel}, \& {Turok}}]{pen94}
{Pen}, U.-L., {Spergel}, D.~N., \& {Turok}, N. 1994, \prd, 49, 692

\bibitem[{{Planck Collaboration} {et~al.}(2016){Planck Collaboration}, {Ade},
  {Aghanim}, {Akrami}, {Aluri}, {Arnaud}, {Ashdown}, {Aumont}, {Baccigalupi},
  {Banday}, \& et~al.}]{pl15}
{Planck Collaboration} {et~al.} 2016, \aap, 594, A16

\bibitem[{{Planck Collaboration} {et~al.}(2020{\natexlab{a}}){Planck
  Collaboration}, {Akrami}, {Ashdown}, {Aumont}, {Baccigalupi}, {Ballardini},
  {Band ay}, {Barreiro}, {Bartolo}, {Basak}, {Benabed}, {Bersanelli},
  {Bielewicz}, {Bond}, {Borrill}, {Bouchet}, {Boulanger}, {Bucher}, {Burigana},
  {Calabrese}, {Cardoso}, {Carron}, {Casaponsa}, {Challinor}, {Colombo},
  {Combet}, {Crill}, {Cuttaia}, {de Bernardis}, {de Rosa}, {de Zotti},
  {Delabrouille}, {Delouis}, {Di Valentino}, {Dickinson}, {Diego}, {Donzelli},
  {Dor{\'e}}, {Ducout}, {Dupac}, {Efstathiou}, {Elsner}, {En{\ss}lin},
  {Eriksen}, {Falgarone}, {Fernandez-Cobos}, {Finelli}, {Forastieri},
  {Frailis}, {Fraisse}, {Franceschi}, {Frolov}, {Galeotta}, {Galli}, {Ganga},
  {G{\'e}nova-Santos}, {Gerbino}, {Ghosh}, {Gonz{\'a}lez-Nuevo}, {G{\'o}rski},
  {Gratton}, {Gruppuso}, {Gudmundsson}, {Hand ley}, {Hansen}, {Helou},
  {Herranz}, {Hildebrandt}, {Huang}, {Jaffe}, {Karakci}, {Keih{\"a}nen},
  {Keskitalo}, {Kiiveri}, {Kim}, {Kisner}, {Krachmalnicoff}, {Kunz},
  {Kurki-Suonio}, {Lagache}, {Lamarre}, {Lasenby}, {Lattanzi}, {Lawrence}, {Le
  Jeune}, {Levrier}, {Liguori}, {Lilje}, {Lindholm}, {L{\'o}pez-Caniego},
  {Lubin}, {Ma}, {Mac{\'\i}as-P{\'e}rez}, {Maggio}, {Maino}, {Mand olesi},
  {Mangilli}, {Marcos-Caballero}, {Maris}, {Martin},
  {Mart{\'\i}nez-Gonz{\'a}lez}, {Matarrese}, {Mauri}, {McEwen}, {Meinhold},
  {Melchiorri}, {Mennella}, {Migliaccio}, {Miville-Desch{\^e}nes}, {Molinari},
  {Moneti}, {Montier}, {Morgante}, {Natoli}, {Oppizzi}, {Pagano}, {Paoletti},
  {Partridge}, {Peel}, {Pettorino}, {Piacentini}, {Polenta}, {Puget}, {Rachen},
  {Reinecke}, {Remazeilles}, {Renzi}, {Rocha}, {Roudier},
  {Rubi{\~n}o-Mart{\'\i}n}, {Ruiz-Granados}, {Salvati}, {Sandri}, {Savelainen},
  {Scott}, {Seljebotn}, {Sirignano}, {Spencer}, {Suur-Uski}, {Tauber},
  {Tavagnacco}, {Tenti}, {Thommesen}, {Toffolatti}, {Tomasi}, {Trombetti},
  {Valiviita}, {Van Tent}, {Vielva}, {Villa}, {Vittorio}, {Wandelt}, {Wehus},
  {Zacchei}, \& {Zonca}}]{planckcomponent}
---. 2020{\natexlab{a}}, \aap, 641, A4

\bibitem[{{Planck Collaboration} {et~al.}(2020{\natexlab{b}}){Planck
  Collaboration}, {Aghanim}, {Akrami}, {Ashdown}, {Aumont}, {Baccigalupi},
  {Ballardini}, {Banday}, {Barreiro}, {Bartolo}, {Basak}, {Battye}, {Benabed},
  {Bernard}, {Bersanelli}, {Bielewicz}, {Bock}, {Bond}, {Borrill}, {Bouchet},
  {Boulanger}, {Bucher}, {Burigana}, {Butler}, {Calabrese}, {Cardoso},
  {Carron}, {Challinor}, {Chiang}, {Chluba}, {Colombo}, {Combet}, {Contreras},
  {Crill}, {Cuttaia}, {de Bernardis}, {de Zotti}, {Delabrouille}, {Delouis},
  {Di Valentino}, {Diego}, {Dor{\'e}}, {Douspis}, {Ducout}, {Dupac}, {Dusini},
  {Efstathiou}, {Elsner}, {En{\ss}lin}, {Eriksen}, {Fantaye}, {Farhang},
  {Fergusson}, {Fernandez-Cobos}, {Finelli}, {Forastieri}, {Frailis},
  {Fraisse}, {Franceschi}, {Frolov}, {Galeotta}, {Galli}, {Ganga},
  {G{\'e}nova-Santos}, {Gerbino}, {Ghosh}, {Gonz{\'a}lez-Nuevo}, {G{\'o}rski},
  {Gratton}, {Gruppuso}, {Gudmundsson}, {Hamann}, {Handley}, {Hansen},
  {Herranz}, {Hildebrandt}, {Hivon}, {Huang}, {Jaffe}, {Jones}, {Karakci},
  {Keih{\"a}nen}, {Keskitalo}, {Kiiveri}, {Kim}, {Kisner}, {Knox},
  {Krachmalnicoff}, {Kunz}, {Kurki-Suonio}, {Lagache}, {Lamarre}, {Lasenby},
  {Lattanzi}, {Lawrence}, {Le Jeune}, {Lemos}, {Lesgourgues}, {Levrier},
  {Lewis}, {Liguori}, {Lilje}, {Lilley}, {Lindholm}, {L{\'o}pez-Caniego},
  {Lubin}, {Ma}, {Mac{\'\i}as-P{\'e}rez}, {Maggio}, {Maino}, {Mandolesi},
  {Mangilli}, {Marcos-Caballero}, {Maris}, {Martin}, {Martinelli},
  {Mart{\'\i}nez-Gonz{\'a}lez}, {Matarrese}, {Mauri}, {McEwen}, {Meinhold},
  {Melchiorri}, {Mennella}, {Migliaccio}, {Millea}, {Mitra},
  {Miville-Desch{\^e}nes}, {Molinari}, {Montier}, {Morgante}, {Moss}, {Natoli},
  {N{\o}rgaard-Nielsen}, {Pagano}, {Paoletti}, {Partridge}, {Patanchon},
  {Peiris}, {Perrotta}, {Pettorino}, {Piacentini}, {Polastri}, {Polenta},
  {Puget}, {Rachen}, {Reinecke}, {Remazeilles}, {Renzi}, {Rocha}, {Rosset},
  {Roudier}, {Rubi{\~n}o-Mart{\'\i}n}, {Ruiz-Granados}, {Salvati}, {Sandri},
  {Savelainen}, {Scott}, {Shellard}, {Sirignano}, {Sirri}, {Spencer},
  {Sunyaev}, {Suur-Uski}, {Tauber}, {Tavagnacco}, {Tenti}, {Toffolatti},
  {Tomasi}, {Trombetti}, {Valenziano}, {Valiviita}, {Van Tent}, {Vibert},
  {Vielva}, {Villa}, {Vittorio}, {Wand elt}, {Wehus}, {White}, {White},
  {Zacchei}, \& {Zonca}}]{pl18}
---. 2020{\natexlab{b}}, \aap, 641, A6

\bibitem[{{Planck Collaboration} {et~al.}(2020{\natexlab{c}}){Planck
  Collaboration}, {Akrami}, {Ashdown}, {Aumont}, {Baccigalupi}, {Ballardini},
  {Band ay}, {Barreiro}, {Bartolo}, {Basak}, {Benabed}, {Bersanelli},
  {Bielewicz}, {Bock}, {Bond}, {Borrill}, {Bouchet}, {Boulanger}, {Bucher},
  {Burigana}, {Butler}, {Calabrese}, {Cardoso}, {Casaponsa}, {Chiang},
  {Colombo}, {Combet}, {Contreras}, {Crill}, {de Bernardis}, {de Zotti},
  {Delabrouille}, {Delouis}, {Di Valentino}, {Diego}, {Dor{\'e}}, {Douspis},
  {Ducout}, {Dupac}, {Efstathiou}, {Elsner}, {En{\ss}lin}, {Eriksen},
  {Fantaye}, {Fernandez-Cobos}, {Finelli}, {Frailis}, {Fraisse}, {Franceschi},
  {Frolov}, {Galeotta}, {Galli}, {Ganga}, {G{\'e}nova-Santos}, {Gerbino},
  {Ghosh}, {Gonz{\'a}lez-Nuevo}, {G{\'o}rski}, {Gruppuso}, {Gudmundsson},
  {Hamann}, {Hand ley}, {Hansen}, {Herranz}, {Hivon}, {Huang}, {Jaffe},
  {Jones}, {Keih{\"a}nen}, {Keskitalo}, {Kiiveri}, {Kim}, {Krachmalnicoff},
  {Kunz}, {Kurki-Suonio}, {Lagache}, {Lamarre}, {Lasenby}, {Lattanzi},
  {Lawrence}, {Le Jeune}, {Levrier}, {Liguori}, {Lilje}, {Lindholm},
  {L{\'o}pez-Caniego}, {Ma}, {Mac{\'\i}as-P{\'e}rez}, {Maggio}, {Maino}, {Mand
  olesi}, {Mangilli}, {Marcos-Caballero}, {Maris}, {Martin},
  {Mart{\'\i}nez-Gonz{\'a}lez}, {Matarrese}, {Mauri}, {McEwen}, {Meinhold},
  {Mennella}, {Migliaccio}, {Miville-Desch{\^e}nes}, {Molinari}, {Moneti},
  {Montier}, {Morgante}, {Moss}, {Natoli}, {Pagano}, {Paoletti}, {Partridge},
  {Perrotta}, {Pettorino}, {Piacentini}, {Polenta}, {Puget}, {Rachen},
  {Reinecke}, {Remazeilles}, {Renzi}, {Rocha}, {Rosset}, {Roudier},
  {Rubi{\~n}o-Mart{\'\i}n}, {Ruiz-Granados}, {Salvati}, {Savelainen}, {Scott},
  {Shellard}, {Sirignano}, {Sunyaev}, {Suur-Uski}, {Tauber}, {Tavagnacco},
  {Tenti}, {Toffolatti}, {Tomasi}, {Trombetti}, {Valenziano}, {Valiviita}, {Van
  Tent}, {Vielva}, {Villa}, {Vittorio}, {Wandelt}, {Wehus}, {Zacchei}, {Zibin},
  \& {Zonca}}]{pl18stat}
---. 2020{\natexlab{c}}, \aap, 641, A7

\bibitem[{{Platen} {et~al.}(2008){Platen}, {van de Weygaert}, \&
  {Jones}}]{pla08}
{Platen}, E., {van de Weygaert}, R., \& {Jones}, B.~J.~T. 2008, \mnras, 387,
  128

\bibitem[{{Rees} \& {Sciama}(1968)}]{ree68}
{Rees}, M.~J., \& {Sciama}, D.~W. 1968, Nature, 217, 511

\bibitem[{Ringeval {et~al.}(2016)Ringeval, Yamauchi, Yokoyama, \&
  Bouchet}]{ringeval2016large}
Ringeval, C., Yamauchi, D., Yokoyama, J., \& Bouchet, F.~R. 2016, Journal of
  Cosmology and Astroparticle Physics, 2016, 033

\bibitem[{{Rudnick} {et~al.}(2007){Rudnick}, {Brown}, \& {Williams}}]{rud07}
{Rudnick}, L., {Brown}, S., \& {Williams}, L.~R. 2007, \apj, 671, 40

\bibitem[{Sachs \& Wolfe(1967)}]{sac67}
Sachs, R.~K., \& Wolfe, A.~M. 1967, Astrophys. J., 147, 73, [Gen. Rel.
  Grav.39,1929(2007)]

\bibitem[{Sch{\"a}fer(2008)}]{schafer2008integrated}
Sch{\"a}fer, B.~M. 2008, Monthly Notices of the Royal Astronomical Society,
  388, 1403

\bibitem[{Smith \& Huterer(2010)}]{smi08}
Smith, K.~M., \& Huterer, D. 2010, Mon. Not. Roy. Astron. Soc., 403, 2

\bibitem[{Szapudi {et~al.}(2015)}]{sza14}
Szapudi, I., {et~al.} 2015, Mon. Not. Roy. Astron. Soc., 450, 288

\bibitem[{{Turok}(1989)}]{tur89}
{Turok}, N. 1989, Physical Review Letters, 63, 2625

\bibitem[{Turok \& Spergel(1990)}]{tur09}
Turok, N., \& Spergel, D. 1990, Phys. Rev. Lett., 64, 2736

\bibitem[{Vielva(2010)}]{vie10}
Vielva, P. 2010, Advances in Astronomy, 2010, 1

\bibitem[{{Vielva} {et~al.}(2004){Vielva}, {Mart{\'{\i}}nez-Gonz{\'a}lez},
  {Barreiro}, {Sanz}, \& {Cay{\'o}n}}]{vie04}
{Vielva}, P., {Mart{\'{\i}}nez-Gonz{\'a}lez}, E., {Barreiro}, R.~B., {Sanz},
  J.~L., \& {Cay{\'o}n}, L. 2004, \apj, 609, 22

\bibitem[{Vilenkin \& Shellard(2000)}]{vil00}
Vilenkin, A., \& Shellard, E. P.~S. 2000, {Cosmic Strings and Other Topological
  Defects} (Cambridge University Press)

\bibitem[{Zhang \& Huterer(2010)}]{zha09}
Zhang, R., \& Huterer, D. 2010, Astropart. Phys., 33, 69

\bibitem[{Zibin(2014)}]{zib14}
Zibin, J.~P. 2014, Comment on "A Supervoid Imprinting the Cold Spot in the
  Cosmic Microwave Background"

\end{thebibliography}
\bibliographystyle{apj}

\end{document}